\def\d{\delta}

\def\G{\Gamma}

\def\bpl{\beta^+}
\def\bmn{\beta^-}

\def\bp{{\beta_{ij}}^+}
\def\bm{{\beta_{ij}}^-}
\def\bb{\bar\beta_{ij}}

\def\ha{{1\over 2}}

\def\be{\begin{equation}}
\def\te{\end{equation}}
\def\bea{\begin{eqnarray}}
\def\nn{\nonumber}
\def\tea{\end{eqnarray}}

\newskip\humongous \humongous=0pt plus 1000pt minus 1000pt
\def\caja{\mathsurround=0pt}
\def\eqalign#1{\,\vcenter{\openup2\jot \caja
        \ialign{\strut \hfil$\displaystyle{##}$&$
        \displaystyle{{}##}$\hfil\crcr#1\crcr}}\,}
\newif\ifdtup

\documentstyle[12pt]{article}

\makeatletter                    
\@addtoreset{equation}{section}  
\makeatother                     

\begin{document}

\title{Quantum Statistical Field Theory in Gravitation and Cosmology}

\author{B. L. Hu\thanks{ Email: hu@umdhep.umd.edu}\\
{\small Department of Physics, University of Maryland,
College Park, MD 20742, USA} }
\date{January 20, 1994}
\maketitle
\centerline{\it Lectures given at the Canadian Summer School for Theoretical
Physics and}
\centerline{\it the Third International Workshop on Thermal Field Theories
and Applications}
\centerline{\it Banff, Canada, Alberta, August 15-28, 1993}
\centerline{(umdpp 94-45) }

\begin{abstract}
We describe how the concepts of quantum open systems and the methods of
closed-time-path (CTP) effective action and influence functional (IF)
can be usefully applied to the analysis of statistical mechanical problems
involving quantum fields in gravitation and cosmology. In the first lecture
we discuss in general terms the relevance of open system concepts in the
description of a variety of physical processes, and outline the basics of the
CTP and IF formalisms. In the second lecture we illustrate the IF method with
a model of two interacting quantum fields, deriving the influence action via a
perturbative expansion involving the closed-time-path Green functions. We show
how noise of quantum fields can be defined and derive a general fluctuation-
dissipation relation for quantum fields. In the third lecture we discuss the
problem of backreaction in semiclassical gravity with the example of a scalar
field in a Bianchi Type-I universe. We show that the CTP effective action not
only yields a real and causal equation of motion with a dissipative term
depicting the effect of particle creation, as was found earlier, it also
contains a noise term measuring  the fluctuations in particle number and
governing the metric fluctuations. The particle creation-backreaction problem
can be understood as a manifestation of a fluctuation-dissipation relation for
quantum fields in dynamic spacetimes, generalizing Sciama's observation
for black hole Hawking radiation.  A more complete description of semiclassical
gravity
is given by way of an Einstein-Langevin equation, the conventional theory
based on the expectation value of the energy momentum tensor being its
mean field approximation. We also mention related problems of interest,
including the dissipative nature of effective field theory and the theory
of fluctuations, instability and phase transition as applied to the problem
of transition from general relativity to quantum gravity.

\end{abstract}

\newpage
\section*{Outline and Summary}

The purpose of these lectures is to develop a quantum field theory suitable for
the description of non-equilibrium quantum systems for problems in gravitation
and cosmology. This is not a review, but only an exposition and summary of
our recent work and some thoughts for future development.

\subsection {Aim of Program}

{\bf Schwinger-Keldysh} On the theoretical side,
I will summarize the relation of the closed-time-path (CTP, or
Schwinger-Keldysh)
\cite{SchKel,ctpChinese,DeWJor,CH87,CH88,CH89}
functional formalism with the influence-functional (IF, or Feyman-Vernon)
\cite{FeyVer,CalLeg83,Gra,HPZ1,HPZ2,HuBelgium} method.
The CTP method is now quite well received and applied to  particle physics
problems
involving thermal fields, as is witnessed in this meeting. After its inception
in the 60's it was used in selected condensed matter problems.
In the first paper Esteban Calzetta and I
wrote on the CTP formalism \cite{CH87} we adapted it to quantum fields in
curved spacetime \cite{BirDav} and applied it to the backreaction problem  of
particle production in the early universe \cite{cpcbkr,HarHu} as the highlight
of
semiclassical gravity theory \cite{scg}.
As we expected, to get a real and causal effective equation of motion
for these systems,
this method was not only neat, but absolutely indispensible. Only after going
through our own initiation process
\footnote{I became aware of this subject rather late, after hearing a talk at
the 1980 Guangzhou Particle Physics Meeting by the authors of the well-known
review \cite{ctpChinese}.
I knew even then that this is the method which can render the complex
geometry obtained from the Schwinger-DeWitt (`in-out') effective action
formalism \cite{SchDeW} real. It is also most suitable for treating
quantum statistical processes in cosmology, which is prevalently
non-equibrium in nature. But I wanted to see the whole picture for myself so
I started from ground level by first learning (and ended up constructing a
good part of) finite temperature quantum field theory in curved, and
especially,
dynamic spacetimes \cite{ftf}. Such was my initiation to thermal field theory.}
had we learned from Jordan's paper that Bryce DeWitt had
been advocating this method for some time \cite{DeWJor}. (We should have
guessed!) Our second paper was the development of a quantum kinetic theory
(in flat spacetime) with the CTP formalism by combining the multiple-source
(Cornwall-Jackiw-Tomboulis \cite{CJT}) formalism for treating correlation
functions and the  Wigner function \cite{Wigner} techniques to derive
the kinetic equations \cite{CH88}. That was our first venture into
nonequilibrium statistical
field theory. An extension to curved spacetime was done with Habib \cite{CHH}.
Paz later applied the CTP formalism to cosmological backreaction problem with
finite temperature fields, including a nice calculation of the
viscosity function for reheating in the inflationary cosmology \cite{Paz90}.

{\bf Feynman-Vernon}
At about the same time Schwinger and Keldysh \cite{SchKel} wrote their
seminal papers  which established the CTP formalism, Feynman and Vernon
\cite{FeyVer} wrote a very important though lesser known paper which
introduced the influence functional formalism. Both were about harmonic
oscillators and Brownian motion.
The influence functional formalism was popularized by the work of
Calderia and Leggett in 1983 \cite{CalLeg83}.
My initiation, or rather, reintroduction to this method
\footnote{ Even though the IF formalism was
discussed in detail in Feynman and Hibbs' classic 1965 book,
I did not pay attention to it in my first reading of that book, because
I read it only for understanding the path integral method and never bothered
with
the statistical mechanics part. I think I am not alone in this kind of
regreted oversight and missed opportunity.}
came from my insistence to understand the true statistical
properties of quantum fields in general \cite{HuPhysica} and to uncover
the missing noise term in the cosmological backreaction problem in particular.
Dissipation in the dynamics of a system arising from particle creation of
a quantum field has been shown by many researchers before \cite{cpcbkr,HarHu},
the last definitive result was obtained
from the calculation via the CTP formalism \cite{CH87}. But the statistical
meaning of dissipation in these contexts was not clear and
the expected accompanying effects of fluctuation and noise were apparently
nowhere to be seen in the CTP formalism. I summarized these inquiries
in the First Thermal Field Workshop and gave my tentative replies to them
in the form of three conjectures \cite{HuPhysica}:\\
1) That colored noise associated with quantum field fluctuations is generally
expected in gravitation and cosmology;\\
2) That the backreaction of particle creation in a dynamical spacetime
can be viewed as the manifestation of a generalized fluctuation-dissipation
relation; and \\
3) That all effective field theories, including semiclassical gravity or
even quantum gravity (to the extent that it could be viewed as an effective
field theory), are intrinsically dissipative in nature.\\
These conjectures are logical consequences of the open system concept
\cite{qos}
applied to quantum fields in curved spacetime. Asking such questions
also led me to a closer examination of the relationship between the influence
functional and the CTP formalisms.
I will discuss these issues and methodology in what follows.
The search and discovery which led to a formulation of
quantum field theory in curved spacetime \cite{BirDav} in terms of
non-equilibrium statistical mechanics are also described in
\cite{HPS,HuWaseda}.\\

On the physical side, the range of problems in gravitation and cosmology
which statistical field theory is particularly useful for
is described in two recent reviews:
Ref. \cite{HuWaseda} discusses quantum statistical processes
in the early universe, with special emphasis on the entropy of quantum fields
and spacetimes; Ref. \cite{HMLA} describes the origin and nature
of noise in quantum fields, and how it is related to
particle creation in black holes and the early unverse.

\subsection{Topics of my Three Lectures at the Summer School}

The titles and contents of my three lectures presented in this summer school
are:

1. Quantum Decoherence and Uncertainty:
The influence functional formalism and application to foundational problems in
quantum mechanics and statistical mechanics.

2. Quantum Origin of Noise and Fluctuations:
 Application to galaxy formation problems in inflationary cosmology.

3. Dissipation, Fluctuation and Backreaction:
 Application to problems in semiclassical gravity and quantum cosmology.

Let me summarize the main points of these lectures. \\

In Lecture 1\\
a) {\it Decoherence}~~
We discussed the issue of decoherence in the problem of quantum to
classical transition using the quantum master equation derived via the
influence functional formalism.  It is based on the work of Hu, Paz and Zhang
on quantum Brownian motion (QBM) in a general environment \cite{HPZ1}. For the
bigger picture, see the recent reviews \cite{decrev} on the consistent or
decoherent history formulation \cite{conhis} and the environment-induced
decoherence program \cite{envdec} of quantum mechanics.\\
b) {\it Uncertainty}~~
We derived the uncertainty principle at finite temperature,
pointing out that under general conditions the decoherence time
is the same as the time when thermal fluctuations overtake quantum
fluctuations.
This is based on the work of Hu and Zhang \cite{HuZhaUncer}.
A review of the issue of decoherence and
uncertainty in terms of quantum and thermal noise is given in my talk at
Third Drexel Workshop on Quantum Dynamics of Chaotic Systems
\cite{HuZhaDrexel}. See also the recent paper on this topic via
information theory by Anderson and Halliwell \cite{AndHal}.

In Lecture 2:\\
a) {\it Quantum Noise}~~ We showed how noise can be identified from the
imaginary part of the influence action, and how its characteristics depends on
the coupling of the system with the environment, the spectral density of
the environment and other factors. We studied a model of the Brownian particle
coupled
nonlinearly with a bath of harmonic oscillators. We deduced the noise
autocorrelation functions and a generalized fluctuation-dissipation relation
for colored and multiplicative noise. The details are contained in the second
QBM paper of Hu, Paz and Zhang \cite{HPZ2}.\\
b) {\it Quantum Fluctuations}~~We suggested a proper way to deduce noise from
quantum fluctuations of the inflaton field in the quantum theory of
galaxy formation, and showed why the conventional way of simply treating the
short wavelength sector of a {\it free} field as the  source of a classical
Langevin equation is flawed \cite{StoInf}. We also examined
the preconditions for a {\it classical} stochastic
equation to emerge from the wave equations of quantum fields, e.g., that the
long wave-length sector would have to decohere for this to be possible. We
illustrated how to tackle these two basic issues with two coupled
$\lambda \phi^4$ fields. In the course of this exposition we also
pointed out how the closed-time-path Green functions enter in the
perturbative calculation of the influence functional.
The details of this problem worked out by Hu, Paz and Zhang are recorded
in my talk at the Chateau du Pont O'ye Meeting on the Origin of Structure in
the Universe \cite{HuBelgium}.

In Lecture 3 \\
a) {\it Backreaction and Dissipation}~~
For semiclassical gravity we discussed the backreaction of particle
production on the dynamics of a spacetime, using the anisotropy dissipation
in a Bianchi Type-I universe as  example. This was done primarily for the
purpose
of illustrating the use of the closed-time-path method  and to point out
how the statistical mechanical meaning of dissipation can indeed be sought with
the quantum open system concept. It was in asking these questions which
led us to the discovery of the connection between the CTP and IF formalisms.
First derived by Calzetta and Hu \cite{CH87,CH89}, the result was
reviewed in my talk at the First Thermal Field Workshop \cite{HuPhysica}.\\
b) {\it Minisuperspace as Open System}~~
For quantum cosmology we discussed the question of the validity of
the minisuperspace approximation \cite{mss},
wherein only the homogeneous cosmologies
are quantized and the inhomogeneous cosmologies ignored \cite{HuErice}.
We used an interacting quantum field model and calculated the effect of
the inhomogeneous modes on the homogeneous mode via the CTP effective action.
This effect manifests in the effective equation of motion for the system as a
dissipative term. For quantum cosmology, this backreaction
turns the Wheeler-De Witt (WDW) \cite{WdW}
equation for the full superspace into an effective
WDW equation for the minisuperspace with dissipation. This is detailed in
the work by Sinha and Hu \cite{SinHu,Sinha,HPS}. I also discussed
the possibility of defining a gravitational entropy based on these concepts
at the Waseda Conference on Quantum Physics and the Universe \cite{HuWaseda}.\\

\subsection {Recent Results}

In the time between the conference and the submission of this report
some new results were obtained by Calzetta, Matacz, Sinha and I. They fall
under the subject matters of Lectures 2 and 3, which I add here as topics 2c
and 3c:

2c) {\it Thermal Radiance}~~
As a model for an open system in the environment of a quantum scalar field,
we studied a Brownian particle interacting with a bath of parametric
oscillators.
The single oscillator can be a particle detector, a mode for the field (e.g.,
the homogeneous inflaton mode), or  the scale factor of the universe.
We derived an expression for the  influence functional in terms
of the Bogolubov coefficients relating the amplitudes of the bath modes
under parametric amplification, which in the second quantized sense
represents particle creation in a dynamical spacetime.  With this result one
can discuss the
statistical mechanical meaning and origin of thermal radiance observed in an
accelerated detector, a black hole and the de Sitter universe, known as
the Unruh and the Hawking effects \cite{Haw,Unr,GibHaw}, in terms of
the excitation of vacuum fluctuations or quantum noise in the quantum field.
This study which sets the stage for the influence functional formalism approach
to
quantum field theory in curved spacetime is detailed in Hu and Matacz
\cite{HM2}.
The nature and effects of quantum noise in gravitation and cosmology is
discussed in my recent talk at the Fluctuation and Order Workshop
at Los Alamos \cite{HMLA}.

3c) {\it Einstein-Langevin Equation and Fluctuation-Dissipation Relation}~~
Calzetta and I  continued our earlier investigation of the backreaction problem
in semiclassical gravity to show how noise and
fluctuations can also be obtained with the CTP formalism in addition
to dissipation and decoherence \cite{nfsg}.
We derived an expression for the CTP effective action in terms of the Bogolubov
coefficients and showed how noise is related to the fluctuations in particle
number. Matacz and I used a cumulant expansion on the influence functional
\cite{HM3}  to extract  the noise associated with the matter field and
to derive an Einstein-Langevin equation as the equation of motion for the
the dynamics of spacetime as an open system. Through these work, we have
extended
the old framework of semiclassical gravity
based on the mean field theory of the Einstein equation with a source driven by
the expectation value of the energy momentum tensor, to that based on a
Langevin equation which describes also the fluctuations of matter fields and
spacetime. The backreaction effect of matter fields on the dynamics of
spacetime
exemplified by the anisotropy damping due to particle creation can be viewed as
a manifestation of the fluctuation-dissipation relation
\cite{fdr,Sciama,Mottola}.
This conjecture of mine
\cite{HuPhysica} inspired by Sciama's observation on Hawking radiation in
black holes  was proven by Sinha and I recently \cite{HuSin}. We show this
result in Part III below.

\subsection{Content of this Report}

Instead of repeating or rewording the already published work,
I prefer to select a few topics in quantum field theory in flat
and curved spacetimes from our current research work  which
carry some representative meaning  and discuss how this new approach via
quantum statistical field theory can provide new insights and directions.
The following topics are covered in this report:\\

Part I: Concepts and Methods in Statistical Field Theories\\
1) Quantum Open Systems: Coarse-Graining and Backreaction\\
2) The Influence Functional Approach to Statistical Field Theory\\
3) The Closed-Time-Path  Functional Formalism in Quantum FIeld Theory\\
4) The Consistent Histories Formulation of Quantum Mechanics\\

Part II: Noise and Fluctuations in Quantum Fields\\
1) Perturbation Theory   \\
2) Noise and Fluctuations   \\
3) Langevin Equation and Fluctuation-Dissipation Relation\\
4) Related Problems\\

Part III: Einstein-Langevin Equation in Semiclassical Gravity\\
1) Backreaction and Dissipation\\
2) Noise\\
3) Einstein-Langevin Equation and Fluctuation-Dissipation Relation\\
4) Related Problems\\

These parts are based on ongoing (unpublished) work with Esteban Calzetta
\cite{nfsg} (Part I), Juan Pablo Paz and Yuhong Zhang
\cite{Zhang,qsf} (Part II) and Sukanya Sinha \cite{HuSin} (Part III).
The reader is referred to these work for further details.
\vskip 1cm
\centerline{\bf Part I}

\section{Concepts and Methods in Statistical Field Theories}

We first give a physical reason for treating familiar quantum processes
with the conceptual framework of open systems. We then
give a brief summary of the relation of the closed-time-path,
the influence functional and the decoherence functional formalisms
used in the path-integral approach to quantum and statistical mechanics
and field theory.

\subsection{Quantum Open Systems: Coarse-Graining and Backreaction}

In physical systems containing many degrees of freedom one often
attempts to select out a small set of variables to render the problem
technically tractable while preserving its essence.   Familiar
examples are abound, e.g., thermodynamics from statistical mechanics,
hydrodynamic limit of kinetic theory, collective dynamics in nuclear
physics \cite{Wilson}. When one starts from the microscopic picture, one
distinguishes
the variables which depict the system of interest from those which can
affect the system but whose detail is otherwise of lesser interest or no
importance. To make a sensible distinction involves recognizing and
devising a set of criteria to separate the relevant from the irrelavant
variables.  This procedure is simplified when the two sets of variables
possess very different characteristic time or length or energy scales or
interaction strengths.
An example is the slow-fast variables separation in the Born-Oppenheimer
approximation in molecular physics where the nuclear variables are assumed
to enter adiabatically as parameters in the electronic wave function.
Similar separation is possible in quantum cosmology between the `heavy'
gravitational sector and the `light' matter sector characterized by the
Planck mass (see, e.g., \cite{GerGru} and references therein).
In statistical physics this separation can be made formally with projection
operator techniques \cite{ProjOp}.
This usually results in a nonlinear integro-differential equation for the
relevant variables, which contains the causal and correlational information
from their interaction with the irrelavant variables.

Apart from finding some way of {\bf separating} the overall closed system
(the ``universe") into a `relevant' part of primary interest (the open system)
and an `irrelevant' part of secondary interest (the environment)
in order to render calculations possible,
one also needs to devise some {\bf averaging} scheme
to reduce and reconstitute the detailed information  of the environment
such that its effect on the system can be traced to some simple macroscopic
functions.
This involves introducing certain {\bf coarse-graining}
measures.  It is by the imposition of such measures that an environment is
turned into a bath, and certain  macroscopic characteristics
such as temperature, chemical potential can be devised for its simple
description. A coarse-grained description of the effect of the environment
on the system (in terms of, say, the transport coefficients in hydrodynamics)
is qualitatively very different from the detailed description (in terms of the
underlying microscopic dynamics).  A familiar example in many body theory
used for simplifying the effect of the environment is the
independent particle model, where, say, a nucleon is assumed to be affected by
all other nucleons only via an averaged two-body potential.  Mean field
approximation in quantum field theory shares the same spirit, where the
effect of quantum fluctuations of fields is described in terms of a
renormalization  of the basic physical parameters
of the system.

How the environment affects the open system is called the {\bf backreaction}
effect.  By referring
to an effect as backreaction, it is implicitly assumed that a system of
interest
is preferentially identified, that one cares much less about the details of the
other sector (the `irrelevant' variables, the `environment', etc).
The backreaction can be significant, but should not be too overpowering, so
as to invalidate the separation scheme.  To what extent one views the
interaction of the two sectors as interaction or as backreaction is
reflexive of the degree one decides to keep or discard the information in
one versus the other. It also depends on their interaction strength.
The behavior of the complete (closed) system requires a self-consistent
solution of equations governing both sectors.  Examples are the Hartree-Fock
approximation in atomic physics or nuclear physics, where the system could
be described by the wave function of the electrons obeying the Schr\"odinger
equations with a potential determined by the charge density of the
electrons themselves via the Poisson equation \cite{TDHF}.
In a cosmological backreaction problem, the system
is a classical spacetime, whose dynamics is determined by Einstein's
equations with sources given by particles produced by the vacuum excited by
the dynamics of spacetime and depicted by wave equations in curved spacetime
\cite{BirDav}. Much of the physics of open systems is concerned with the
practicality and validity of these procedures. They are,
1) the identification and separation of the physically interesting
variables which make up the open system -- oftentimes one needs to come up with
the appropriate collective variables, 2) the `averaging' away of the
environment or irrelevant variables -- how different coarse-graining measures
affect the final result is important,  and 3) the evaluation of the averaged
effect of the environment on the system of interest.
We will refer to these procedures as {\it separation, coarse-graining}
and {\it backreaction} for short.

These considerations surrounding an open system are common and essential
not only to well-posed and well-studied examples of many-body systems
like molecular, nuclear and condensed-matter physics, they also
bear on some basic
issues at the foundations of quantum mechanics and statistical mechanics,
such as decoherence and the existence of classical limit \cite{GelHar2},
the emergence of time and spacetime from quantum gravity \cite{HarLH}.
Our recent work is concerned with the general problem of how these
considerations can be applied to both classical and quantum
gravitational systems.  This includes problems
in Einstein's classical theory of general relativity,
quantum field theory in curved
spacetime (semi-classical gravity) and quantum gravity, as well as their
related problems in relativistic cosmology and quantum cosmology.
Gravitational systems are of interest not only because of the special
characteristics of gravity, but also because
they dwell on extreme conditions such as that which exist in the black holes
and the early universe, often pushing the physical laws to their limits.
Inquiries on the quantum and statistical properties of spacetime and fields
raise many new issues in fundamental physics,
whose resolution can provide interesting new insights into the nature of
physical laws and the structure of the universe \cite{HK,MG5,HuSpain}.\\

\centerline{* ~~~ * ~~~ *}

In terms of methodology, two formalisms have been used effectively
for the description of quantum open systems, especially pertaining to
the coarse-graining and backreaction problems:
the closed time path effective action (CTP, or Schwinger-Keldysh) formalism
\cite{SchKel,ctpChinese,DeWJor,CH87,CH88,CH89} for obtaining a real and  causal
 equation of motion with dissipation;
and the influence functional (IF, or Feynman-Vernon
\cite{FeyVer,CalLeg83,Gra,HPZ1,HPZ2,HuBelgium})
formalism for identifying the noise and fluctuations in the environment.
We give here a brief description of these formalisms and their interconnection.
We also sketch the decoherent history formulation of quantum mechanics
\cite{conhis}
as we will use this conceptual framework to apply  the IF and CTP formalisms
to the analysis of semiclassical gravity theory. The following summary is
adapted from \cite{nfsg}.

\subsection{The Influence Functional Approach to Statistical Field Theory}

The IF approach \cite{FeyVer} is designed to deal with a situation in which the
system $S$ described, say, by the $x$ fields is interacting with
an environment $E$, described by the $q$ fields.
The full quantum system is described by a density matrix
$\rho (x,q;x',q',t)$. If we are only interested in
the state of the system as influenced by the overall effect,
but not the precise state, of the environment,
then the reduced density matrix $\rho_r(x,x',t)=\int~dq~\rho (x,q;x',q,t)$
would provide the relevant information. (The subscript $r$ stands for
reduced.) It is propagated in time from $t_i$ by the propagator ${\cal J}_r$:

$$
\rho_r(x,x',t)
=\int\limits_{-\infty}^{+\infty}dx_i\int\limits_{-\infty}^{+\infty}dx'_i~
 {\cal J}_r(x,x',t~|~x_i,x'_i,t_i)~\rho_r(x_i,x'_i,t_i~)
\label{pathint}
$$

Assuming that the action
of the coupled system decomposes as $S=S_s[x]+S_e[q]+S_{int}[x,q]$,
and that the initial density matrix factorizes (i.e.,
takes the tensor product form),
$\rho (x,q;x',q',t_i)=\rho_s(x,x',t_i)\rho_e(q,q',t_i)$, the
propagator for the reduced density
matrix is given by

$$
{\cal J}_r(x,x',t|~x_i,x'_i,t_i)=
 \int_{x_i}^{x_f} Dx~ \int_{x'_i}^{x'_f} Dx '~e^{iS_{eff}[x, x', t]}
$$
where
\be
S_{eff}[x, x', t] \equiv S_s[x]-S_s[x']+S_{IF}[x,x',t] \label{Jr}
\te
is the full effective action and $S_{IF}$ is the influence action.
(They are called ${\cal A}$ and $\delta {\cal A}$ in \cite{HPZ1} )
The influence functional $\cal F$ is defined as

\be
{\cal F}[x, x', t] \equiv e^{iS_{IF}[x,x',t]}\equiv\int~dq_f~dq_i~ dq_i'~
  \int_{q_i}^{q_f}Dq~\int_{q'_i}^{q_f}Dq'~
e^{i(S_e[q]+S_{int}[x ,q]-S_e[q']-S_{int}[x ',q'])} \rho_e(q_i,q'_i,t_i).
\label{SIF}
\te
$S_{IF}$ is typically
complex; its real part ${\cal R}$, containing the dissipation kernel
$ \hat \mu$, contributes to the renormalization of $S_s$, and
yields the dissipative terms in the effective equations of motion.
The imaginary part ${\cal I}$,
containing the noise kernel $\hat \nu $, provides the information about
the fluctuations induced on the system through its coupling to the environment.
Since
the connection between these kernels and their effect on the physical processes
of dissipation and fluctuation has been discussed at lenght elsewhere
(cfr. Ref.  \cite{HPZ1}), we shall  limit
ourselves here only to a schematic summary. \footnote{This
simplified schematic discussion is really just for the illustration of main
ideas, not for precision or completeness. The reader is referred to
\cite{HPZ1,HPZ2,HuBelgium,HMLA,HM2} for details on the discussion of the
process of decoherence in quantum to classical transition,
the origin and nature of quantum noise, the fluctuation-dissipation relation
and the explicit derivations of the master, Fokker-Planck and Langevin
equations
for the models of a Brownian particle in a general environment and interacting
quantum fields in  cosmological spacetimes.}

The main  features  of  the  influence  action
follow from  the  elementary  properties  $S_{IF}(x,x')=-S_{IF}(x',x)^*$  and
$S_{IF}(x,x)=0$, which can  be  deduced from its definition,
and derived in the final analysis  from  the  unitarity  of the underlying
quantum theory of the closed system.
If we decompose $S_{IF}$ in its real and imaginary parts,
 $S_{IF}={\cal R}+i{\cal I}$, then
${\cal R}(x,x')=-{\cal R}(x',x)$,
${\cal I}(x,x')={\cal I}(x',x)$, and ${\cal R}(x,x)={\cal I}(x,x)=0$.
Keeping only
quadratic terms, we may write

\be
S_{IF}(x,x')=\int~dt~dt'~\{ {1\over 2}(x-x')(t)\hat \mu (t,t')(x+x')(t')
+{i\over 2}(x-x')(t)\hat \nu (t,t')(x-x')(t')\}
\label{SIFquad}
\te
where $\hat \mu $ and $\hat \nu $ stand for the real dissipation and
noise kernels respectively ($\hat \mu \equiv 2\eta$, $\hat \nu \equiv 2\nu$,
in the notations of \cite{HPZ1}).
It is convenient to express $\hat \mu $ as
$\hat \mu (t,t')=-\partial_{t'}\hat \gamma (t,t')$, and rewrite

\be
S_{IF}(x,x')=\int~dt~dt'~\{ {1\over 2}(x-x')(t)\hat \gamma (t,t')(\dot x+\dot
x')
(t')+{i\over 2}(x-x')(t)\hat \nu (t,t')(x-x')(t')\}
\te
The physical meaning of  the $\hat \gamma$ kernel may be elucidated by deriving
the mean field equation of motion for the mean value of the system variable
$\bar x$. It is

\be
{\delta S_s\over\delta\bar x(t)}+\int~dt'~\hat \gamma (t,t')
{d\bar x(t')\over dt'}=0
\te
The term  containing $\hat \gamma$
represents the  backreaction  of  the  environment  on  the
system.  It causes the dissipation  of energy from the system by an amount
(integrated over the whole history of the system)

\be
\Delta E=\int ~dt~dt'~\hat \gamma (t,t')\dot {\bar x}(t)\dot {\bar x}(t').
\te
Thus we see that the even part of the kernel $\hat \gamma$ is associated with
dissipation, while the odd part can be assimilated to a
nondissipative environment-induced change in the system dynamics.
In quantum field-theoretic applications, the odd part of $\hat \gamma$ will
contain formally infinite terms which can  be  absorbed  in the classical
action for the system via standard renormalization procedures
\cite{BirDav}. For  simplicity, we shall assume
that only the even part of  $\hat \gamma$  is  left after renormalization has
been
carried out.

In  general, the $\hat \gamma$ and $\hat \nu $ kernels are nonlocal;
however,  their
main  features    are  manifest  already  under  the  local  approximation
$\hat \gamma\sim\hat \gamma_0\delta (t-t')$,
$\hat \nu \sim \hat \nu_0\delta (t-t')$.  The influence
action then takes the form

\be
S_{IF}(x,x')=\int~dt~\{ {1\over 2}(x-x')(t)\hat \gamma_0(\dot x+\dot x')
(t)+{i\over 2}(x-x')(t)\hat \nu_0(x-x')(t)\}
\te

Assuming an action functional of
the  simple  form  $S_s[x]\sim\int\{\ha    \dot    x^2-V(x)\}$,    it   is
straightforward  to derive the master  equation  for  the  reduced  density
matrix \cite{FeyVer,CalLeg83}

\be
i{\partial\rho_r\over\partial t}\sim
\{ [-\ha \partial_x^2+V(x)]-[-\ha \partial_{x'}^2+V(x')]
-i{\hat \gamma_0\over 2}(x-x')[{\partial\over\partial x}-{\partial\over\partial
x'}]
-i {\hat \nu_0\over 2} (x-x')^2\}\rho_r
\label{master}
\te

The object `closest' (see \cite{cor}) to a classical distribution function
is the Wigner function \cite{Wigner}

\be
f_W(X,p)=\int~dy~e^{ipy}\rho_r(X+{y\over 2},X-{y\over 2})
\label{defwigner}
\te
where $X \equiv (1/2)(x + x'), y \equiv x - x'$.
The master equation (\ref{master}) implies (to lowest order in a
Kramers-Moyal expansion) the Fokker-Planck equation
\cite{Chandra,vK}

\be
\{{\partial\over\partial t}+p{\partial\over\partial X}-V'{\partial\over
\partial p}\}f_W=(\hat \gamma_0 {\partial\over\partial p} p+
{\hat \nu_0\over 2}{\partial^2\over\partial p^2})f_W
\label{Fokker}
\te
(where $V'=dV/dx$).
{}From this equation one can see clearly the stochasticity in the semiclassical
dynamics. The Fokker-Planck equation admits the equilibrium solution

\be
f^{eq}_W\sim e^{-(2\hat \gamma_0 /\hat \nu_0)[(p^2/2)+V(x)]}
\te
which depicts a finite-temperature stationary state for the system.
In particular, a fluctuation-dissipation theorem
$\hat \nu _0=2\hat \gamma_0 \langle p^2\rangle_{eq}$
can be easily derived. If the environment acts as a heat bath, then $
\langle p^2\rangle_{eq}\sim k_BT$, and this reduces to the
Einstein-Kubo formula for the dispersion coefficient.

\subsection{The Closed-Time-Path Functional Formalism in
Quantum Field Theory}

In the CTP approach, our goal is not to follow the dynamics of the
full density matrix, or even the system part, but only the expectation
values of the fields as they unfold in time. This evolution is
governed by a real and causal equation of motion, which is obtained
from the CTP effective action by a variational principle.

Let $\psi$ be the quantum fields in the theory, and $\bar\psi$ their
expectation
values for any given initial state. Consider pairs of histories
$(\psi,\psi')$ defined on all spacetime, with the property that $\psi(T^0)
=\psi'(T^0)$ for a given very large time $T^0$ (in practice, $T^0\to +\infty$).
Assume for simplicity
(more general choices are also possible \cite{CH88}) that the fields
were originally in their vacuum state $\vert 0in\rangle$
and that the bath fields $\psi, \psi'$
are linearly coupled to external sources $J, J'$.
A closed time
path is the path which runs from the initial field configuration
$ \psi_i(\vec x) $ at time $ s=-\infty $ to the final field configuration
$ \psi_f(\vec x) $ at time $ s=+\infty $ through the positive time branch
linked to $J$,
then returns to the field configuration $ \psi'_i(\vec x) $ at time
$ s=-\infty $ through the negative time branch linked to $J'$.
The CTP generating functional for the $ \psi $ field over the vacuum state
is given by

\bea
Z[J,J']~=~e^{iW[J,J']}~
& =\int d\psi_f(\vec x)~
   \Bigl<0in\Bigl|\tilde T\exp\biggl\{
   -{i\over\hbar}\int\limits_{-\infty}^{+\infty}ds\int d\vec x
   J'(x)\hat\psi(x)\biggr\}\Bigr|\psi_f\Bigr> \cr
& ~~~~~~ \times
   \Bigl<\psi_f\Bigl|T\exp\biggl\{
   {i\over\hbar}\int\limits_{-\infty}^{+\infty}ds\int d\vec x
   J(x)\hat\psi(x)\biggr\}\Bigr|0in\Bigr>              \label{Wcan}
\tea
where $ T $ denotes temporal order, $ \tilde T $ denotes
anti-temporal order and $ \psi_0(\vec x) $ is the field configuration
corresponding to the vacuum state of the $ \psi $ field.
Observe  that  the generating  functional  $W$  is  totally  defined
once the {\it in}  state  $\vert 0in\rangle$ is chosen
and that $W\equiv 0$ whenever $J=J'$.
Now introduce the path integral representation

\bea
Z[J, J']
& =\int d\psi_f(\vec x)
   \int\limits_{\psi_i(\vec x)}^{\psi_f(\vec x)}d\psi
   \int\limits_{\psi'_i(\vec x)}^{\psi_f(\vec x)}d\psi'
   \exp {i\over\hbar}\biggl\{ S[\psi]
    +\int\limits_{-\infty}^{+\infty}ds\int d^3\vec x~
    J(x)\psi(x) \cr
& ~~~~~   -S[\psi']
   -\int\limits_{-\infty}^{+\infty}ds\int d^3\vec x~
    J'(x)\psi'(x)\biggr\}
\tea
The expectation values can be obtained as

\be
\bar\psi={\d W\over\d J}, ~~~
\bar\psi'=-{\d W\over\d J'}
\label{eqmotion}
\te
The physically relevant situation under consideration corresponds to
setting $J=J'=0$.

The full CTP effective action is just the Legendre transform of $W$

\be
\G_{CTP} [\bar\psi,\bar\psi']=W[J,J']-J\bar\psi+J'\bar\psi'
\te
where now the sources are thought of as functionals of the background fields
$\bar\psi,\bar\psi'$. In particular, the equations of motion are the inverses
of
Eqs. (\ref{eqmotion})

\be
{\d \G_{CTP}\over\d \bar\psi}=-J, ~~~
{\d \G_{CTP}\over\d \bar\psi'}=J'
\te
The physical situations correspond to solutions of the homogeneous equations
at $\bar\psi =\bar\psi'$. These equations are real and causal. Moreover,
$\G_{CTP} [\bar\psi,\bar\psi']
=-\G_{CTP}^*[\bar\psi',\bar\psi]$, and $\G_{CTP} [\bar\psi,\bar\psi]\equiv 0$.
As  the  generating  functional  itself, the CTP effective action is  totally
defined once the initial quantum state is given.

To apply this formalism to the situation above, we should substitute
the $\psi$ field by the pair $(x,q)$. When the physical situation requires
treating the $x$ and $q$ fields asymmetrically,
as is the case when, say, only the system field $x$ is relevant,
we do not couple the $q$ field to an external source. (In a perturbative
evaluation of the CTP generating functional, this means discarding
all graphs with $q$ fields on some external leg.)
Comparing the path integral expression for the generating functional
with the IF approach described earlier (\ref{Jr}), we find

\be
e^{iW[J,J']}=\int~Dx~ Dx'~
{}~e^{i(S_s[x]-S_s[x']+Jx-J'x'+S_{IF}[x,x',+\infty ])}
\te
Conversely, we may describe the full IF effective action $S_{eff}$ as the
full CTP effective action
\be
\Gamma_{CTP}[x, x'] = S_s [x] - S_s [x'] + \Gamma[x, x']
\te
for the quantum $q$ fields interacting with external c-number
$x$ fields specialized to the expectation values of its arguments.

In the semiclassical approximation, one can
neglect Feynman graphs containing closed $x$ field loops corresponding to
quantum effects of the $x$ fields.
Then the path integral and the Legendre transformation may be
computed explicitly, yielding

\be
\G_{CTP} [x,x']\approx S_{eff}[x,x',+\infty ]
\label{GSIF}
\te
This equation shows the equivalence between the (full) CTP effective action and
the (full) IF effective action, or $\Gamma$ and $S_{IF}$.
{}From this one may derive the semiclassical
equations of motion for the expectation values of the $x$ field. We
see that the noise kernel does not contribute to these equations,
because, it being even under the exchange of $x$ and $x'$, its
variation vanishes at the coincidence point. However, as we shall see below,
and is also clear from the master equation point of view \cite{HPZ1},
the noise kernel determines the dynamics governing the deviations
from the expectation value.

\subsection{The Consistent Histories Formulation of Quantum Mechanics}

Let us  now  relate these concepts and techniques in
statistical field theory to the quantum to classical transition
problem via the consistent histories formulation of quantum
mechanics \cite{conhis,GelHar2,PazSin}.

In the consistent  or decoherent histories approach, the  complete description
of a coupled $x,q$ system is given in terms of fine-grained histories
$x(t),q(t)$. These histories are quantum in nature, i. e. it is possible
in principle to observe
interference effects between
different generic histories. A classical description is
acceptable only at the level of coarse-grained histories, and to the
extent that interference effects between these histories become
unobservable. Let us adopt the simple coarse-graining procedure
of leaving the $q$ field unspecified. Then each coarse-grained
history is labelled by a possible evolution of the $x$ field,
and the interference effects between histories are measured
by the decoherence functional (DF)

\be
{{\cal D}}[x,x']=
{}~e^{i(S_s[x]-S_s[x'])}
\int dq_i~dq'_i ~dq_f~\int Dq~~ Dq'~
e^{i(S_e[q]+S_{int}[x,q]-S_e[q']-S_{int}[x',q'])}
\rho_e(q_i,q'_i,t_i)
\te
which  is  the  fundamental  object  of  the theory. (For a more formal
definition see \cite{conhis}.)
The coarse-grained histories $x(t)$ can be described classically if
and only if the decoherence functional is approximately diagonal, that is,
${\cal D}[x,x'] \simeq 0$ whenever $x\not= x'$.  The conditions leading to this
in quantum mechanics is the focus of many current studies, to which we refer
the readers for the details. For quantum cosmology the issue is complicated
by the problem of time, and there even the definition of the decoherence
functional can be ambiguous \cite{HarLH}. In the problem of transition
from quantum cosmology  to semiclassical gravity, a WKB time is usually
assumed. Using the minisuperspace quantum cosmology of Bianchi Type-I universe
\cite{mss} as example, Paz and Sinha \cite{PazSin}
showed that an influence functional appears naturally from a reduced
density matrix by tracing out the matter fields. They discussed the decoherence
between WKB branches of the wave function and tried to relate it to the
notion of decoherence between spacetime histories. In our discussion of
semiclassical gravity in Part III, we  will assume
that this essential task can be accomplished in some satisfactory way,
from which one can write down the decoherent functional \cite{PazSin,nfsg}.

\be
{\cal D}[x,x']=
e^{i(S_s[x]-S_s[x']+S_{IF}[x,x',\infty])}=
e^{i\G_{CTP} [x,x']}
\te
Already at this formal level notice that decoherence can occur only when
the noise kernel is nonzero, which signals the presence
of spontaneous fluctuations in the system.

For an observer confined (by
necessity or by choice) to the level of coarse grained descriptions,
dynamical evolution must be described in terms of mutually exclusive
histories, all interference
effects having been suppressed below the accuracy of his observation devices.
For example, if he chooses to describe the evolution of the system in terms
of its Wigner Function $f_W$, he will
now interpret it as an actual ensemble average, describing the
joint  evolution  of  the    bundle    of    coarse-grained   histories.
Correspondingly,
he will regard Eq. (\ref{Fokker}) as a classical Fokker-Planck
equation. Now the classical random process
described  by  Eq.   (\ref{Fokker})  is  not  deterministic;    rather,  it
describes the evolution of an ensemble of particles whose individual orbits
obey the Langevin-type equations

\be
\dot x=p~~~~~;~~~~\dot p=-V'-\hat \gamma p+ \xi
\label{Langevin}
\te
where $\xi$ represents a noise term with autocorrelation
$\langle \xi (t)\xi (t')\rangle = \hat\nu(t,t')$. (The ordinarily assummed
gaussian and white nature of the noise follows only from
a quadratic and local noise kernel, which describes rather special cases
in cosmological
situations, see \cite{HuBelgium,HMLA}). Thus, the observer confined to
a coarse-grained history will  conclude that
semiclassical evolution is stochastic.
Note that the statistical properties of this random evolution
are totally determined by
the  decoherence  functional; no {\it ad hoc} assumptions
on the behavior of quantum fluctuations are necessary.

To see how noise arises from the IF formalism, one can rewrite the part in the
influence action containing the noise kernel as

\be
e^{-\ha \int~dt~(x-x') \hat \nu (x-x')}\equiv\int~D\xi~e^{i\int~dt~\xi(x-x')}
e^{-\ha \int~dt~\xi {\hat\nu}^{-1}\xi}
\label{Foutrans}
\te
Therefore the action of the environment on the system may be described by
adding the external source term $-\int~x\xi$ to the system action $S_s$,
and averaging over external sources with the proper weight
\cite{Zhang,HPZ2,HM2}. Variation of this effective action directly yields
the Langevin equations (\ref{Langevin}).
This is how noise can be understood
as a stochastic force from the environment acting on the system.

In Part II  we will apply these formalisms to discuss noise and
fluctuations from interacting quantum fields in flat space. Then in Part III
we will discuss the problem of dissipation and fluctuation in a curved
spacetime setting.
\vskip 1cm
\centerline{Part II}
\section{Noise and Fluctuations in Quantum Fields}

Consider two independent self-interacting scalar fields in Minkowsky spacetime:
$ \phi(x) $ depicting the system, and $ \psi(x) $ depicting the bath.
The classical action for these two fields are given respectively by:
\be
S[\phi]
=\int d^4x~
  [{1\over 2}\partial_{\nu}\phi(x)\partial^{\nu}\phi(x) - V(\phi)]
=S_0[\phi]+S_I[\phi]     \label{Sphi}
\te
\be
S[\psi]
=\int d^4x~
 [ {1\over 2}\partial_{\mu}\psi(x)\partial^{\mu}\psi(x) - V(\psi)]
=S_0[\psi]+S_I[\psi]       \label{Spsi}
\te
where $V[\phi], V[\psi]$ are the self-interaction potentials. For
a $\phi^4$ interaction,
\be
V[\phi] = {1\over 2}m^2_{\phi}\phi^2(x) + {1\over 4!}\lambda_{\phi}\phi^4(x),
\te
and similarly for $V[\psi]$. Here, $ m_{\phi} $ and $ m_{\psi} $ are
the bare masses and $ \lambda_{\phi} $ and $ \lambda_{\psi} $ are the bare
self-coupling constants  for the $ \phi(x) $ and $ \psi(x) $ fields
respectively.
In (\ref{Spsi}) we have written $S[\psi]$ in terms of a free part $S_0$ and
an interacting part $S_I$ which contains $ \lambda_{\psi}$.
Assume these two scalar fields interact via a polynomial coupling
of the form
\be
S_{int} =\int d^4x~\Bigl\{ v[\phi(x)] \psi^k(x) \Bigr\} \label{Sint}
\te
where $ v[\phi(x)] \equiv -\lambda_{\phi\psi} f[\phi(x)]$ is
the vertex function with coupling constant $ \lambda_{\phi\psi} $,
which we assume to be small and of the same order as
$ \lambda_{\phi} $, $ \lambda_{\psi} $.

The total classical action of the combined system is
\be
S[\phi,\psi]
=S[\phi]+S[\psi]+S_{int}[\phi,\psi]
\te

\noindent The total density matrix of the combined system plus bath field is
defined by
\be
\rho[\phi,\psi,\phi',\psi',t]
=<\phi,\psi|~\hat\rho(t)~|\phi',\psi'>
\te
\noindent where $ |\phi> $ and $ |\psi> $ are the eigenstates of the field
operators $ \hat\phi(x) $ and $ \hat\psi(x) $, namely,
\be
\hat\phi(\vec x) |\phi>=\phi(\vec x) |\phi>, ~~~
\hat\psi(\vec x) |\psi> = \psi(\vec x) |\psi>
\te
%
%

Since we are primarily interested in the behavior of the system, and of
the environment only to the extent in
how it influences the system, the object of interest
is the reduced density matrix defined by
\be
\rho_{red}[\phi,\phi',t]=\int d\psi \rho[\phi,\psi,\phi',\psi,t]
\te
%
For technical convenience, let us assume that the total density matrix
at an initial time is factorized, i.e.,
that the system and bath are statistically independent,
\be
\hat\rho(t_i)
=\hat\rho_{\phi}(t_i)\times\hat\rho_{\psi}(t_i)
\te
\noindent where $ \hat\rho_r(t_i) $ and $ \hat\rho_{\psi}(t_i) $ are the
initial
density matrix operator of the $ \phi $ and $ \psi $ field respectively,
the former being equal to the reduced density matrix $\hat \rho_r$ at $t_i$
by this assumption.
The reduced density matrix of the system field $ \phi(x) $ evolves
in time following
\be
\rho_r[\phi_f,\phi'_f,t]
=\int d\phi_i\int d\phi'_i~
 {\cal J}_r[\phi_f,\phi'_f,t~|~\phi_i,\phi'_i,t_i]~
 \rho_r[\phi_i,\phi'_i,t_i]
\te

\noindent where ${\cal J}_r$ is the evolutionary operator
of the reduced density matrix:
\be
{\cal J}_r[\phi_f,\phi'_f,t~|~\phi_i,\phi'_i,t_i]
=\int\limits_{\phi_i(\vec x)}^{\phi_f(\vec x)}D\phi
 \int\limits_{\phi'_i(\vec x)}^{\phi'_f(\vec x)}D\phi'~
 \exp {i\over \hbar} S_{eff}[\phi, \phi']
\te
where
\be
 S_{eff} [\phi, \phi'] \equiv S[\phi]-S[\phi'] + S_{IF}[\phi, \phi']
                                    \label{infact}
\te
is the full IF effective action and $S_{IF}$ is the influence action.
The influence functional $ {\cal F}[\phi,\phi']~ $ is defined as
\be
\eqalign{
{\cal F}[\phi,\phi']= & e^{{i\over\hbar} S_{IF}[\phi, \phi']} =
   \int d\psi_f(\vec x)
   \int d\psi_i(\vec x)
   \int d\psi'_i(\vec x)~
   \rho_{\psi}[\psi_i,\psi'_i,t_i]~
   \int\limits_{\psi_i(\vec x)}^{\psi_f(\vec x)}D\psi
   \int\limits_{\psi'_i(\vec x)}^{\psi_f(\vec x)}D\psi' \cr
&  \times\exp {i\over\hbar}\Bigl\{S[\psi]+S_{int}[\phi,\psi]
    -S[\psi']-S_{int}[\phi',\psi'] \Bigr\} \cr }   
\te
which summarizes the averaged effect of the bath on the system.
We have seen from Sec. 1.3 that for a zero-temperature bath
(i.e., the environment field $ \psi $ is in a vacuum state,
$\hat\rho_b(t_i)=|0><0| $),
the influence functional ${\cal F}$ is formally equivalent to the CTP
vacuum generating functional,
and the influence action $S_{IF}$ in (\ref{infact}) is the usual CTP
vacuum effective action.

\subsection{Perturbation Theory}

The above is the formal framework we shall work with. Let us now develop
a perturbation theory for evaluating the influence action. If
$ \lambda_{\phi\psi} $ and $ \lambda_{\psi} $ are assumed to be small
parameters,
the influence functional can be calculated perturbatively
by making a power expansion of $ \exp {i\over\hbar}\bigl\{S_{int}+S_I\bigr\} $.
Up to second order in $~\lambda $,
and first order in $\hbar$ (one-loop), the influence action is given by
\be
\eqalign{
S_{IF}[\phi,\phi']
& =~~\biggl\{
   <S_{int}[\phi,\psi]>_0
  -<S_{int}[\phi',\psi']>_0 \biggr\} \cr
& +{i\over 2\hbar}\biggl\{
   <\Bigl[S_{int}[\phi,\psi]\Bigr]^2>_0
  -\Bigl[<S_{int}[\phi,\psi]>_0\Bigr]^2 \biggr\} \cr
& -~{i\over \hbar}\biggl\{<S_{int}[\phi,\psi]S_{int}[\phi',\psi']>_0
   -<S_{int}[\phi,\psi]>_0
    <S_{int}[\phi',\psi']>_0 \biggr\} \cr
& +{i\over 2\hbar}\biggl\{
    <\Bigl[S_{int}[\phi',\psi']\Bigr]^2>_0
   -\Bigl[<S_{int}[\phi',\psi']>_0\Bigr]^2 \biggr\} \cr }
\te
\noindent where the quantum average of a physical variable $Q[\psi, \psi']$
over the unperturbed action $ S_0[\psi] $ is defined by
\be
\eqalign{
<Q[\psi,\psi']>_0
& =\int d\psi_f(\vec x)\int d\psi_i(\vec x)\int d\psi'_i(\vec x)~
    \rho_{\psi}[\psi_i,\psi'_i,t_i]  \cr
& \times
    \int\limits_{\psi_i(\vec x)}^{\psi_f(\vec x)} D\psi
    \int\limits_{\psi'_i(\vec x)}^{\psi_f(\vec x)} D\psi'~
    \exp {i\over\hbar}\Bigl\{ S_0[\psi]-S_0[\psi'] \Bigr\}
     \times Q[\psi,\psi'] \cr
& \equiv Q\Bigl[{\hbar \delta \over i\delta J(x)},~
    -{\hbar \delta \over i\delta J'(x)} \Bigr]~
    {\cal F}^{(0)}[J,J']~\biggl|_{J=J'=0} \cr }
\te

\noindent Here, ${\cal F}^{(0)}[J,J'] $ is the
influence functional of the free bath field, assuming a linear coupling with
external sources $J$ and $J'$:
\bea
{\cal F}^{(0)}[J,J']
&=\int d\psi_f(\vec x)
  \int d\psi_i(\vec x)
  \int d\psi'_i(\vec x)~
  \rho_{\psi}[\psi_i,\psi'_i,t_i]
  \int\limits_{\psi_i(\vec x)}^{\psi_f(\vec x)} D\psi
  \int\limits_{\psi'_i(\vec x)}^{\psi_f(\vec x)} D\psi' \\
& \times\exp {i\over\hbar}\Bigl\{ S_0[\psi]+\int d^4x J(x)\psi(x)
    -S_0[\psi']-\int d^4x J'(x)\psi'(x) \Bigr\}   \label{infnal}
\tea
Let us define the following free propagators of the $ \psi $ field
\be
<\psi(x)\psi(y)>_0=iG_{++}(x,y)
\te
\be
<\psi'(x)\psi'(y)>_0=-iG_{--}(x,y)
\te
\be
<\psi(x)\psi'(y)>_0=-iG_{+-}(x,y)
\te
We see that these are just the familiar Feynman, Dyson and positive-frequency
Wightman propagators of a free scalar field given respectively by
\be
G_{++}(x,y)=G_F(x-y)
=\int{d^np\over (2\pi)^n}e^{ip(x-y)}
 {1\over p^2-m^2_{\psi}+i\epsilon}
\te
\be
G_{--}(x,y)=G_D(x-y)
=\int{d^np\over (2\pi)^n}e^{ip(x-y)}
 {1\over p^2-m^2_{\psi}-i\epsilon}
\te
\be
G_{+-}(x,y)=G^+(x-y)
=\int{d^np\over (2\pi)^2}e^{ip(x-y)}2\pi
i\delta(p^2-m^2_{\psi})\theta(p^0)
\te
The perturbation calculation by means of Feynman diagrams
for $\lambda \phi^4$ theory in the CTP formalism
has been carried out before for quantum fluctuations \cite{CH87}
and for coarsed-grained fields \cite{cgea,SinHu}.
For bi-quadratic coupling,

\be
S_{int}[\phi,\psi] =\int d^4x~\Bigl\{
 -\lambda_{\phi\psi}\phi^2(x)\psi^2(x) \Bigr\}
\te
the influence action
up to the second order in $\lambda$ is given by (cf \cite{HPZ2})
\be
\eqalign{
S_{IF}[\phi,\phi']
& =\int d^4x\Bigl\{-\lambda_{\phi\psi}
   iG_{++}(x,x)\phi^2(x)\Bigl\} -\int d^4x\Bigl\{-\lambda_{\phi\psi}
   iG_{++}(x,x)\phi'^2(x)\Bigr\} \cr
& +\int d^4x\int d^4y~\lambda^2_{\phi\psi}\phi^2(x)~
   \Bigl\{-iG^2_{++}(x,y)\Bigr\}~\phi^2(y) \cr
& -2\int d^4x\int d^4y~\lambda^2_{\phi\psi}\phi^2(x)~
   \Bigl\{-iG^2_{+-}(x,y)\Bigr\}~\phi'^2(y) \cr
& +\int d^4x\int d^4y\lambda^2_{\phi\psi}\phi'^2(x)~
   \Bigl[-iG^2_{--}(x,y)\Bigr\}~\phi'^2(y) \cr }
\te

For a general polynomial-type coupling with $S_{int}$ given by (\ref{Sint}),
the renormalized full effective action to second order in $\lambda$ is given by
\cite{Zhang,HuBelgium}
\be
\eqalign{
S_{eff}[\phi,\phi']
& =\Bigl\{S[\phi]+\delta S_1[\phi]+\delta S_2[\phi]\Bigr\}
  -\Bigl\{S[\phi']+\delta S_1[\phi']+\delta S_2[\phi']\Bigr\}
  +S_{IF}[\phi,\phi'] \cr
& =S_{ren}[\phi]+\int d^4x\int d^4y~{1\over 2}
   \lambda^2_{\phi\psi}f[\phi(x)]  \Delta V^{(k)}(x-y)f[\phi(y)] \cr
& -S_{ren}[\phi']-\int d^4x\int d^4y~{1\over 2}
   \lambda^2_{\phi\psi}f[\phi'(x)] \Delta V^{(k)}(x-y)f[\phi'(y)]\cr
& -{1\over\hbar}\int\limits_{t_i}^tds_x\int d^3\vec x
   \int\limits_{t_i}^{s_y}ds_y\int d^3\vec y~  \lambda^2_{\phi\psi}
\Bigl[f[\phi(x)]-f[\phi'(x)]\Bigr] \mu^{(k)}(x-y)
\Bigl[f[\phi(y)]+f[\phi'(y)]\Bigr] \cr
& +{i\over\hbar}\int\limits_{t_i}^tds_x\int d^3\vec x
   \int\limits_{t_i}^{s_x}ds_y\int d^3\vec y~   \lambda^2_{\phi\psi}
\Bigl[f[\phi(x)]-f[\phi'(x)]\Bigr] \nu^{(k)}(x-y)
\Bigl[f[\phi(y)]-f[\phi'(y)]\Bigr] \cr }
\te
where the subscript on $S$ denotes the order of $\lambda$ in the perturbative
expansion on the action.
Here $ S_{ren}[\phi] $ is the renormalized action of the $\phi $
field,
now with physical mass $ m^2_{\phi r} $ and physical coupling constant
$ \lambda_{\phi r} $, namely,
\be
S_{ren}[\phi]
=\int d^4x\Bigl\{{1\over 2}\partial_{\mu}\phi\partial^{\mu}\phi
-{1\over 2}m^2_{\phi r}\phi^2-{1\over 4!}\lambda_{\phi r}\phi^4\Bigr\}
\te

For the bi-quadratic interaction case
the potential renormalization is
\be
\Delta V^{(2)}(x-y) =\kappa^{(2)}(x-y)-sgn(s_x-s_y)\mu^{(2)}(x-y)
\te
which is symmetric; and
$\mu^{(2)}, \nu^{(2)}$ and $\kappa^{(2)}$
are real nonlocal kernels
\be
\mu^{(2)}(x-y)
={1\over 16\pi^2}\int {d^4p\over (2\pi)^4}~e^{ip(x-y)}
{}~\pi\sqrt{1-{4m^2_{\psi}\over p^2}}~\theta(p^2-4m^2_{\psi})
 \times isgn(p_0)
\te
\be
\nu^{(2)}(x-y)
={2\over 16\pi^2}\int{d^4p\over (2\pi)^4}~e^{ip(x-y)}
{}~\pi~\sqrt{1-{4m^2_{\psi}\over p^2}}~\theta(p^2-4m^2_{\psi})
\te
\be
\kappa^{(2)}(x-y)
=-{2\over 16\pi^2}\int{d^4p\over (2\pi)^4}~e^{ip(x-y)}
 \int\limits_0^1d\alpha\ln\Bigl|1-i\epsilon-
 \alpha(1-\alpha){p^2\over m^2_{\psi}}\Bigr|
\te
Renormalization of the potential which arises
from the contribution of the bath appears only for even $k$ couplings.
For the case $k=1$ the above result is exact.
This is a generalization of the
result obtained in \cite{HPZ2,HuBelgium} where it was shown that the
non-local kernel $\mu^{(k)}(s_1-s_2)$ is associated with dissipation or
the generalized viscosity function that appears in the corresponding
Langevin equation and $\nu^{(k)}(s_1-s_2)$ is associated with the
time correlation function of the stochastic noise term.
In general $\nu$ is nonlocal, which gives rise to colored noises.
Only at high temperatures would the noise kernel become a delta function,
which corresponds to a white noise source. Let us elaborate on the meaning
of the noise kernel.

\subsection{Noise and Fluctuations}

The real part of the influence functional comes from the imaginary part
of the influence action which contains the noise kernel. It is  given by
\be
\exp \Bigl\{ -{1\over\hbar}
\int\limits_{t_i}^tds_x\int d^3\vec x
\int\limits_{t_i}^{s_x}ds_y\int d^3\vec y~  
\Bigl[f[\phi(x)]-f[\phi'(x)]\Bigr] \nu^{(k)}(x-y)
\Bigl[f[\phi(y)]-f[\phi'(y)]\Bigr]  \Bigr\}
\te
where $\nu^{(k)}$ is redefined by absorbing the $\lambda^2_{\phi\psi}$.
This term can be rewritten using a functional Gaussian identity \cite{FeyVer}
to be:
\be
\int D \xi^{(k)}(y) {\cal P}[\xi^{(k)}]\exp\{ {i\over \hbar}\int_0^y
d^4x \xi^{(k)}(x)[f[\phi(x)] - f[\phi'(x)]]\}
\te
where
\be
{\cal P}[\xi^{(k)}] = P^{(k)} \exp\{-{1\over{2 \hbar}}
\int d^4y \int d^4x \xi^{(k)}(x) [\nu^{(k)} (x-y)]^{-1}\xi^{(k)}(y) \}
\te
is the functional distribution of $\xi^{(k)}(s)$ and $P^{(k)}$ is a
normalization factor given by
\be
[P^{(k)}]^{-1} = \int D\xi^{(k)} \exp\{-{1\over \hbar}
\int d^4y \int d^4x \xi^{(k)}(x)[\nu^{(k)}(x-y)]^{-1} \xi^{(k)}(y)\}.
\te

The terms in the effective action describing the coupling between
the noise $ \xi(x) $ and the system field $ \phi(x) $  is then
\be
-\int d^4x~\xi^{(k)}(x) ~\Bigl\{ f[\phi(x)]- f[\phi(x')]~\Bigr\}
\te
In this way the reduced density matrix can be rewritten as
\be
\rho_r(\phi, \phi') = \int D \xi^{(k)}(x){\cal P}[\xi^{(k)}]
\rho_r(\phi,\phi',[\xi^{(k)}]).
\te
Therefore we can view $\xi^{(k)}(x)$ as a
nonlinear external stochastic force and the
reduced density matrix is calculated by taking a stochastic average
over the distribution ${\cal P}[\xi^{(k)}]$ of this source.

Since  the expansion of the action is to the quadratic order,
the associated noise is Gaussian. It is completely characterized by
\bea
{\langle\xi^{(k)}(x)\rangle}_\xi &=& 0 \nn\\
{\langle\xi^{(k)}(x)\xi^{(k)}(y)\rangle}_\xi &=& \hbar\nu^{(k)}(x-y).
\tea
We see that the non-local noise kernel $\nu^{(k)}(x-y)$ is just the two
point auto-correlation function of the external stochastic source
$\xi^{(k)}(x)$ (multiplied by $\hbar$).

In this framework, the expectation value of any functional operator
$Q[\phi]$ of the field $\phi$ is then given by
\bea
\langle Q[\phi]\rangle & = & \int D\xi^{(k)}(x){\cal P}[\xi^{(k)}]\int
d\phi \rho_r(\phi,\phi,[\xi^{(k)}])Q[\phi] \nn \\
& = & {\left\langle {\langle Q[\phi]\rangle}_{quantum}\right\rangle}_{noise}.
\tea
This provides the physical interpretation of $\nu^{(k)}(x-y)$ as a noise or
fluctuation kernel of the quantum field.

\subsection{Langevin Equation and Fluctuation-Dissipation Relation}

We will now derive the semiclassical equation of motion generated
by the influence action $S_{IF}$.
Define a ``center-of-mass" function  $\Sigma$ and a
``relative" function $\Delta$ as follows
\bea
\bar\phi(x) & = & {1\over 2}[\phi(x) + \phi'(x)] \nn\\
\Delta(x) & = & \phi(x) - \phi'(x).
\tea
The equation of motion for $\bar\phi$ is derived by demanding
(cf. \cite{CH87})
\be
 \frac{\delta S_{eff}}{\delta\Delta}\bigg|_{\Delta=0}
=\frac{\delta }{\delta\Delta}\Bigl[S[\phi]-S[\phi']+ S_{IF}[\phi,\phi']\Bigr]
 \bigg|_{\Delta=0}=0.
\te
which gives
\be
-\frac{\partial L_r}{\partial \bar\phi}+\frac{d}{dt}\frac{\partial L_r}
{\partial \dot{\bar\phi}}+2\frac{\partial f(\bar\phi)}{\partial \bar\phi}
\int\limits_0^x d^4y~\gamma^{(k)}(x-y)\frac{\partial f(\bar\phi(y))}{\partial
y}
= F_{\xi}^{(k)}(x)
\te
We see that this is in the form of
a Langevin equation with a nonlinear stochastic force
\be
F_\xi^{(k)}(x) = \xi^{(k)}(s) \frac{\partial f(\bar\phi)}{\partial \bar\phi}.
\te
This corresponds to a multiplicative noise arising from a
nonlinear field coupling (additive if  $f(\bar\phi)=\bar\phi$ ).
$L_{ren}$ is the renormalized effective Lagrangian of
the system  action $S_{eff}$.
The nonlocal kernel $\gamma^{(k)}(t-s)$ defined by
\be
\frac{\partial}{\partial (s_x-s_y)} \gamma^{(k)}(x-y)=\mu^{(k)}(x-y).
\te
is responsible for nonlocal dissipation. Interaction
with the environment field imparts a dissipative force in the effective
dynamics
of the system field given by
\be
F_{ \gamma}^{(k)}(x) =
2\Biggl\{\int d^4y~\mu(x-y)f[\bar\phi(y)]~\Biggr\}\frac{\partial
f(\bar\phi(x))}
{\partial \bar\phi}
\te
Only in special cases like a high temperature ohmic environment when
this kernel becomes a delta function will the dissipation become local.

In the biquadratic coupling example the corresponding stochastic force is
\be
F_{\xi}^{(2)}(x)\sim\xi^{(2)}(x)\bar\phi(x)
\te
The dissipation kernel is
\be
\gamma^{(2)}(x-y)
={1\over 16\pi^2}\int {d^4p\over (2\pi)^4}~e^{ip(x-y)}
 \pi\sqrt{1-{4m^2_{\psi}\over p^2}}~
 \theta(p^2-4m^2_{\psi})~{1\over |p_0|}
\te
and the dissipative force is
\be
F_{\gamma}^{(2)}(x)\sim
\Biggl\{\int d^4y~\mu(x-y)\bar\phi^2(y)~\Biggr\}\bar\phi(x)
\te

As discussed in  \cite{HPZ2},
we can show that a general fluctuation-dissipation relation exists between
the ($l$th order) dissipation and the ($k$th order coupling) noise kernels
in the form

\be
\nu^{(k)}_l(x)=\int d^4y~K^{(k)}_l(x-y)\gamma^{(k)}_l(y)
\te
where the kernel
\be
K^{(k)}_l(x-y)  =\delta^3(\vec x-\vec y)
   \int\limits_0^{+\infty}{d\omega\over\pi} L^{(k)}_l(z)
   l\omega\cos l\omega(s_x-s_y)
\te
where $z \equiv coth \ha \beta \hbar\omega$, and  $L^{(k)}_l$ are temperature-
dependent factors whose explicit expressions can be found in \cite{HPZ2}.
(The linear coupling has $L=z$.)
Apart from the delta function $ \delta^3(\vec x-\vec x') $, the
convolution or fluctuation-dissipation kernel $K$
for quantum fields has exactly the same form as for
the quantum Brownian harmonic oscillator. In general it is a rather complicated
expression, but simplifies at high and zero temperatures.  At high
temperatures,
\be
K^{(k)}_l (s) = {2 k_B T \over \hbar} \delta (s),  ~~~(s \equiv x-y)
\te
this gives back the famous Einstein relation. At zero temperature,
\be
K^{(k)}_l (s) = \int _0^{+\infty} \frac {d \omega}{\pi} \omega \cos \omega s
\te
this is the same as the linear coupling case.
Both limiting forms are independent of $k, l$. In other words, at both high
and zero temperatures, the FDR is insensitive to the way the system is
coupled to the environment. As we have emphasized  earlier,
the  fluctuation-dissipation relation should be understood in a sense more
general than what is usually conceived (under more restrictive conditions
such as near-equilibrium or in the linear response regime).
We view it as a general categorical relation
depicting the stochastic stimulation of the environment and the averaged
response of the system \cite{HPZ2}.

Thus we have given a first-principle derivation of noise from the
quantum fluctuations of interacting quantum fields.
Noise in finite temperature fields are discussed in
\cite{qsf,HM2,HM3} for Minkowski, Robertson-Walker and de Sitter spacetimes.

\subsection{Related Problems}

The above formalism can be applied to a number of problems in statistical
field theory, quantum field theory, gravitation and cosmology.
I'll just mention a few sample problems here but defer the details to
later reports .

1) Application to the problem of {\it galaxy formation from quantum field
fluctuations} in inflationary cosmology. This involves at least two aspects:
a) The origin and nature of noise in quantum fields. Colored noise from
biquadratic coupling between two interacting fields was used as an illustration
of this issue in \cite{Zhang,HuBelgium}. Our discussion here is an extension
of the results of \cite{HuBelgium} to polynomial coupling for quantum fields
following the structure of \cite{HPZ2}.
b) Derivation of a classical Langevin equation from the wave equations of
quantum fields for the description of  gravitational and scalar
perturbations. This involves a study of the decoherence of the wave functions
for certain sector of the spectrum (e.g., long-wavelength modes), which
requires a derivation of the master
equation for inflaton modes in de Sitter universe and an analysis of the
evolution of the diffusion coefficients. The master
equation  for inflaton fields in de Sitter universe was derived in \cite{Zhang}
and a preliminary analysis of the decoherence problem was carried out
in \cite{HuBelgium}. For other recent work on this topic, see e.g.,
\cite{decinf}.

2) Application of non-equilibrium quantum field theory to
the study of {\it phase transitions in the early universe}.
There are at least three aspects in this problem:
a) the field theory aspect, which involves an infrared analysis
of the  effective action; b) the spacetime aspect, where the effects of
spacetime curvature and topology enter; and c) the statistical mechanical
aspect,
where thermal or statistical field theory is invoked. The finite temperature
field theory aspect is well-known to this audience.
The spacetime aspect was studied primarily by researchers in
quantum field theory in curved spacetime. The first stage of work between
1980-86 was described in my 1986 review talk at the Fourth Marcel Grossmann
Meeting \cite{MG4}. The work of O'Connor, Stephens and I on finite size effect
in phase transtions in product spacetimes \cite{HuOC86,OSH}
has as subcases the imaginary-time finite temperature theory,
where, for example, the Coleman-Jackiw-Tomboulis method \cite{CJT}
was used to treat
the infrared behavior of quantum fields under rather general conditions .
For a more recent review on this topic, see, for example, O'Connor and Stephens
\cite{OCSrg}. On the third aspect, the CTP effective action
approach to the non-equilibrium dynamics of phase transition, esp. the quantum
field theory of spinodal decomposition was first studied by Calzetta
\cite{CalSD} based on the relativistic kinetic field theory developed by
Calzetta and Hu \cite{CH88}. One can use the results presented here
for quantum and thermal noise to study noise-induced transitions \cite{Lef},
which is a topic of great interest in early universe phase transitions.
More recently this field has been enriched by
the work of Boyanovsky, de Vega, Gleiser and their coworkers \cite{BoyGle}.

3) The third problem relates to some {\it foundational issues of stochastic
and statistical mechanics}. Specifically, in a field-theory context,
how can one show that  thermal or finite temperature field theory
is the equilibrium limit of statistical fields from the interaction
of a system  with a stochastic bath. This formal question has been
a central issue at the foundation of statistical mechanics
for some time (see, e.g., \cite{Sphon}).
It  arises in two major paradigms of non-equilibrium statistical
mechanics: kinetic theory, and quantum Brownian motion.
Calzetta and I discussed the kinetic model in the 1988 paper \cite{CH88},
the result of which can be used for a  study of this issue, at least for a
special class of field interactions.  For the latter
model, the results of Paz, Zhang and I \cite{HPZ1,HPZ2,HuBelgium}
via the IF formalism can also be used as a start for this non-trivial
problem. From the results of \cite{HPZ1} we can see that many factors
enter into consideration, e.g., the nature of coupling between the system and
the bath, the temperature and the spectral density of the bath.
A more ambitious
program would be to find the conditions which lead to a thermalization of
interacting quantum fields, with no {\it a prioi} assumption of the existence
of
a thermal bath. This is similar to seeking the thermodynamic
limit of a quantum kinetic gas under different ranges and strengths
of interaction.
Calzetta and I have made some preliminary investigation in connection with
decoherence via correlation histories \cite{CHDCH}.
See also a recent paper by Gleiser and Ramos on the issue of
the approach to equilibrium \cite{GleRam}.

4) {\it Dissipative nature of effective field theories}
The formalism above can also be used to probe into the statistical properties
of effective field theories. Effective field theory deals with the appearance
of
one field system in interaction with another when there is a significant
discrepancy in their energy or length scales. The two field theories could
represent the high and low frequency ranges,
the slow and fast regimes, or two sectors of different coupling strengths
in the same theory. If one is only interested
in the low-energy behavior of the complete theory, the effective theory
obtained by `integrating out' the high energy sector will provide most of
the relevant information.  Examples are the Fermi four-point theory
of weak interaction as an effective theory of the Weinberg-Salam
electroweak interaction,  four-dimensional general relativity as an
effective theory of the ten or eleven-dimensional Kaluza-Klein theory of
unified interactions, or Einstein's theory as induced gravity \cite{Adler}.
The interesting issues include the renormalizability of
the effective theory (see, e.g., \cite{Lapage}), the gauge hierarchy problem
and the decoupling theorem (see, e.g., \cite{AppCar,Weinberg,OvrSch}).
Investigation of effective field theory in particle physics has a rather long
history, and the properties are quite well-known.

What we are interested in is the statistical
properties of effective field theories --
this is a topic which has rarely been explored before.
When we calculate the effective action by `intregrating out' one sector
of the theory, 
we don't usually think of their contributions as constituting a noise nor
the resulting dymanics of the effective system as acquiring a dissipative
nature.
This is because in particle physics we are usually interested in the
scattering matrix or transition-amplitude of some process, but not necessarily
the equation of motion; or,  in the case of renormalization,
the ultraviolet behavior of the theory, but not necessarily the dynamics of
fluctuations as in critical phenomena studies. The latter questions are usually
asked in statistical mechanics. Indeed, as we explained in the Introduction to
the closed-time-path formalism in \cite{CH87}, the `in-out' or Schwinger-DeWitt
effective action \cite{SchDeW}
is suitable for transition amplitude and particle creation calculations,
but one needs the in-in or the Schwinger-Keldysh effective action \cite{SchKel}
for deriving a causal equation of motion incorporating the  backreaction
of created particles and other quantum field effects.
In taking the quantum open system point of view and examining the structure
of the influence functional we see that `integrating out' something does not
mean that something simply disappears.
They may `disappear in sight', in a matter of speaking,
from an observer in the subsystem, if he only observes the mean-value of the
variables (source is averaged over the noise distribution).
But traces are left in the noise autocorrelation function as fluctuations
in the source, and in the dissipative component of the effective
equation of motion. The magnitudes  of these fluctuation and
dissipation terms are usually perceived as negligible at the energy scale
of the effective theory. But they are there, they are calculable,
and in priniciple detectable. They can also become important
as the scale moves off the interaction range of the effective theory.
The growth of fluctuations  signals
the instability of the low-energy `effective' phase and possible onset
of phase transition to the next phase (of higher energy). In addition
to the `degree of renormalizability' as a measure of the effectiveness of
a theory (as compared to a `fundamental' theory-- which may never be reachable)
the fluctuations in the environment fields and the dissipation in the
system field dynamics may offer some  useful measure of the
degree of `incompleteness' of an effective field theory,
some hints of the high energy behavior, or signs of `compositeness' of
the underlying theory.
This statistical property of effective theory was suggested earlier
in  \cite{HuPhysica}. Zhang and I have used a coarse-grained effective
action \cite{cgea} for two interacting fields like that discussed here
to derive a renormalization group equation for the coupling constants of the
effective field. Following  the results given here via the influence
functional formalism
we can derive the noise and fluctuations explicitly and study their
physical implications. It is of
interest to expound the significance of these
new statistical effects in relation to the well-known field-theoretical
results. This work is in progress \cite{eft}.
\vskip 1cm
\centerline{\bf Part III}
\section{Einstein-Langevin Equation in Semiclassical Gravity}

After presenting the basic formalism of quantum open systems with the
example of two interacting fields, we now give an example of how it
can be applied to the backreaction problem in semiclassical gravity.
Backreaction effect in curved spacetime encompasses a wide range of problems,
from vacuum fluctuations and polarization effects such as particle creation and
Casimir effect to the renormalization of the energy-momentum tensor
leading to the trace anomaly and the effect of its vacuum expectation
value on the dynamics of spacetime. It is the epitomy of semiclassical
gravity. Our last major investigation of this problem was  done in 1987
with the CTP treatment, leading to a real and causal equation of motion
\cite{CH87}. It was an important advance from the `in-out' (Schwinger-DeWitt
\cite{SchDeW}) to the `in-in' (Schwinger-Keldysh \cite{SchKel}) formalism,
but as we have pointed out before \cite{HuPhysica},
there was an open question about the missing noise and fluctuation terms
which should come hand-in-hand with the dissipation
term engendering anisotropy damping, say, in the Bianchi Type-I example.
As conjectured there, the semiclassical Einstein equation should be in the
form of a Langevin equation, with a colored noise, and the backreaction
problem can be an embodiment of a fluctuation-dissipation relation for
quantum fields in dynamical spacetimes. In a recent paper using the
quantum open system concept and exploiting the formal relation of the CTP
with the influence functional technique,  Sinha and I have indeed
succeeded in showing all three aspects. The following discussion is based on
\cite{HuSin}.

\subsection{Backreaction and Dissipation}

In this example the open system is that of a Bianchi Type-I universe,
and the environment is a massless quantum scalar field.
The  Bianchi type-I metric is given by \cite{Mis67}
\be
ds^2 = a^2 ( d\eta^2 - e^{2\beta}_{ij} dx^i dx^j )
\te
where $a$ is the scale factor,   $\beta_{ij} (\eta)$
is the symmetric and traceless anisotropy matrix and $\eta = \int dt/a$
is the conformal time (a prime denotes a derivative with respect to the
conformal time).
We  will make the assumption that the anisotropy is small and that the
action will be  expanded perturbatively in orders of $\beta$ up to quadratic
order.  To zeroth order in $\beta$ our problem reduces to that of a
spatially flat RW universe with a conformal quantum field.
We will write an enlarged version of the classical action to include the
trace anomaly terms. The full gravitational action expanded out to
second order in $\beta$ is given by \cite{HarHu}
\bea
S_g  &=& \int d\eta \left \{ -6 \kappa {a'}^2 + {[180 {(4\pi)}^2]}^{-1}
\left [ {({a'\over a})}^4  - 3{({a''\over a})}^2 \right ]\right \} \nn \\
     &+& \int d \eta\quad Tr \left \{ \kappa a^2 {\beta'}^2  +
{[180 {(4\pi)}^2 ]}^{-1}\left [ 3 {\epsilon}^{-1} {\beta ''}^2
+ 3 ln(\mu a) {\beta''}^2 - [ ({a''\over a}) {\beta'}^2
- {\beta''}^2] \right] \right \} \nn \\
&+& O(\epsilon)
\tea
Here $\kappa = {(16\pi G)}^{-1}$.
The first line represents the action to zeroth order in $\beta$, i.e,
the action for the flat RW metric including the trace anomaly terms.
We have added the generic $R^2$ terms
in the classical action
to cancel the anticipated ultraviolet divergence and used the standard
counterterms \cite{BirDav} to renormalize the gravitational coupling constants.
$\mu$ is a renormalization constant with units of mass. For simplicity
we assume that we can choose such a renormalization point where the
renormalized cosmological constant and the renormalized coefficients
of the $ R^2 , R_{\mu \nu} R^{\mu \nu}$ and $ R_{\mu \nu \rho \sigma}
R^{\mu \nu \rho \sigma}$ are zero. In accordance with the dimensional
regularization scheme the action was written in an $n$- dimensional
form and expanded in powers of $ \epsilon = n-4$. The singular
term proportional to ${1/ \epsilon}$ appears explicitly in the
action is to be cancelled by anticipated quantum corrections.

The  scalar field action is given by
\be
S_f = \ha \int d^n x {(-g)}^{1\over 2} \left[ g^{\mu \nu}{\partial}_{\mu}
\phi {\partial}_{\nu}\phi - {n-2\over 4(n-1)}R{\phi}^2 \right] .
\te
It can also be expanded up to quadratic order in $\beta$ as above.
Thus our system  variables will be $(a , \beta)$ , and the observer will
follow the dynamics of these, ignoring or being unaware of the details of the
scalar field $\phi$, which plays the role of the environment.

Our main object of interest is the decoherence
functional (DF) between different  histories $ {\cal D} [a^{+}(\eta),
{{\beta}_{ij}}^{+}(\eta); a^{-}(\eta), {{\beta}_{ij}}^{-}(\eta)] $ ,
assuming $\phi$ is completely coarse grained.
Of course, this is a specific kind of coarse graining chosen among the
many possibilities allowed by the decoherence functional. The decoherence
functional for this situation  is given by
\be
{\cal D}[ a^{+},
{\beta}^{+}; a^{-},
{\beta}^{-}] = \int D {\phi}^{+} D{\phi}^{-}
e^{i{(S_{g}[a^{+}, {\beta}^{+}] - S_{g}[a^{-}, {\beta}^{-}]
+ S_{f}[ a^{+}, {\beta}^{+}, {\phi}^{+}] -
S_{f}[ a^{-}, {\beta}^{-}, {\phi}^{-}])}}
\te
In the above expression the histories are assumed to match at some point
$\eta = T$ in the far future and the integration is over field histories
such that ${\phi}^{+}(T) = {\phi}^{-}(T)$. In addition, we must specify
boundary conditions in the distant  past to make the path integral well
defined. We shall assume that $\bpl, \bmn \rightarrow  0$ as $\eta
\rightarrow  -\infty$ , i.e, the anisotropy vanishes in the distant past
and the initial state of the scalar field is the conformal vacuum.
Thus there is a common ``in" particle model for both $+$ and $-$
evolutions defined through the Cauchy data of the conformal particle
model (cf.\cite{nfsg}).

Thus the decoherence functional can be written as
\bea
{\cal D}[ a^{+},
{\beta}^{+}; a^{-},
{\beta}^{-}] &=& e^{i{(S_{g}[a^{+}, {\beta}^{+}] - S_{g}[a^{-}, {\beta}^{-}]
+ S_{IF} [ a^{+},
{\beta}^{+}; a^{-},
{\beta}^{-}])}} \nn \\
& = & e^{i S_{eff} (a^{+}, \label{decfl}
{\beta}^{+}; a^{-},
{\beta}^{-})}
\tea
where $ S_{IF} = \Gamma [ a^{+}, {\beta}^{+}; a^{-}, {\beta}^{-}]$
is the influence or effective action for the scalar field where
$(a, \beta)$ is treated as a classical external field \cite{PazSin,nfsg}.
The decoherence functional
will  be utilized here for a twofold purpose. On the one hand  it can
be used to analyze the extent to which the two histories decohere
based on the smallness of the off-diagonal elements of ${\cal D}$. On the other
hand we will use it to generate the effective equations of motion for
the geometry from the variation of the effective action.
The  regularized influence action  for this problem can be calculated
using Feynman diagram \cite{CH87} or other techniques \cite{PazSin}.
It can be written as
\be
\Gamma[ g^+ , g^-]= \hat\Gamma[g^+, g^-] + i \tilde\Gamma[g^+, g^-]
\te
where $g$ collectively denotes $(a,\beta)$ and both $\hat{\Gamma}$ and
$\tilde{\Gamma}$ are real (called ${\cal R, I}$ in \cite{nfsg}).
Further, $\hat{\Gamma} = {\Gamma}_{div} + {\Gamma}_{ren}$ , where
${\Gamma}_{div}$ and ${\Gamma}_{ren}$ represent the divergent (proportional
to $1/ \epsilon$) and finite contribution to the phase respectively.
${\Gamma}_{div}$ is given by
\be
{\Gamma}_{div} = \int d{\eta}_1 d {\eta}_2
({{\beta}_{ij}}^{+} - {{\beta}_{ij}}^{-} )({\eta}_1)
{\gamma}_{div} 2({\eta}_1 - {\eta}_2)
({{\beta}^{ij}}^{+} - {{\beta}^{ij}}^{-} )({\eta}_2)
\te
where
\be
{\gamma}_{div}({\eta}_1 - {\eta}_2)
= \int_{-\infty}^{+\infty}{d\omega\over 2\pi} e^{i \omega ({\eta}_1 - {\eta}_2)
}
\left[ {-{\omega}^4\over 4 {(4\pi)}^2 (n^2 -1)} {1\over \epsilon}\right],
\te
which exactly cancels the pole in the classical action $S_g$
when the classical action is inserted in the expression for the decoherence
functional.

The rest of the exponent is finite: ${\Gamma}_{ren}$ and $\tilde{\Gamma}$ are
given by
\be
{\Gamma}_{ren} = \int d{\eta}_1 d {\eta}_2
({{\beta}_{ij}}^{+} - {{\beta}_{ij}}^{-} )({\eta}_1)
{\gamma}_{ren}({\eta}_1 - {\eta}_2)
({{\beta}^{ij}}^{+} + {{\beta}^{ij}}^{-} )({\eta}_2)
\te
and
\be
\tilde{\Gamma} = \int d{\eta}_1 d {\eta}_2
({{\beta}_{ij}}^{+} - {{\beta}_{ij}}^{-} )({\eta}_1)
\tilde{\gamma}({\eta}_1 - {\eta}_2)
({{\beta}^{ij}}^{+} - {{\beta}^{ij}}^{-} )({\eta}_2) ,
\te
where the kernels $\gamma_{ren}$ and $\tilde\gamma$ are
\be
\gamma_{ren}(\eta)= -{1\over{60(4\pi)^2}}
\label{odev}
\int_{-\infty}^{+\infty} {{d\omega}\over{2\pi}} ~{\rm e}^{i\omega \eta}
{}~\omega^4 ~
\log(i{{(\omega-i\epsilon)}\over{\mu}})
\te
and
\be
\tilde\gamma(\eta) = ~{1\over{60(4\pi)^2}}
                               \int_{0}^{+\infty} \label{eventilda}
            {{d\omega}\over{2\pi}}{\pi\omega^4} \cos \omega \eta .
\te
Notice that the kernel $\tilde\gamma(\eta)$ is even whereas
$ \gamma_{ren}(\eta)$ contains an odd and even part
given by
\be
\gamma_{odd}(\eta) = ~{1\over{60(4\pi)^2}}
                               \int_{0}^{+\infty} \label{odd}
            {{d\omega}\over{2\pi}}{\pi\omega^4}\sin\omega \eta
\te
and
\be
\gamma_{even}(\eta) =  -{1\over{60(4\pi)^2}}
\int_{-\infty}^{+\infty} {{d\omega}\over{2\pi}} ~\omega^4 \cos
\omega \eta {\rm ln} {|\omega|\over \mu} \label{even}
\te
The kernel $ \gamma_{ren}(\eta)$ is manifestly real and can also be
seen to be causal \cite{CH87}.

Note that $\tilde{\Gamma}$ and ${\Gamma}_{ren}$ play distinct roles here.
$\tilde{\Gamma}$ is responsible for the decoherence between alternative
histories $\bpl$ and $\bmn$ in the sense that it suppresses the
contribution of widely differing histories (or off-diagonal elements)
to the decoherence functional. This feature and its connection to
particle production was explored before in \cite{PazSin,nfsg}.
On the other hand, when
we attempt to derive the effective equation of motion for $\beta$ by
varying the effective action $S_{eff}$,
only ${\Gamma}_{ren}$ contributes  to generating the equation
of motion. The equation of motion obtained under such variation is
identical to the real, causal dissipative equation for $\beta$ obtained
by Calzetta and Hu in \cite{CH87}. In fact, as we will show more
explicitly later, ${\Gamma}_{ren}$ provides the dissipative
contribution to the equation of motion.
Thus in the present form of the decoherence functional ${\Gamma}_{ren}$
contributes only to the equation of motion and not to decoherence, and
$\tilde{\Gamma}$ contributes only to decoherence, and not to the
equation of motion.
However, in the following we will show how $\tilde{\Gamma}$ also plays
the dual role of generating noise and will indeed contribute to
the effective equations of motion with a stochastic source.

Putting these together the renormalized effective action $\Gamma$ or
influence action $S_{IF}$
can be rewritten as
\bea
\Gamma (\bpl,\bmn) & = & ~\int^{\eta}\int^{\eta}
d{\eta}_1d{\eta}_2~{\beta_{ij}}^{+} ({\eta}_1)\Delta V({\eta}_1
-{\eta}_2){\beta^{ij}}^{+}({\eta}_2)\nn \\
& - & ~\int^{\eta}\int^{\eta}
d{\eta}_1 d{\eta}_2~{\beta_{ij}}^{-} ({\eta}_1)\Delta V({\eta}_1
-{\eta}_2){\beta^{ij}}^{-}({\eta}_2)\nn \\
& - & ~\int\limits_0^{\eta}d{\eta}_1\int\limits_0^{{\eta}_1} d{\eta}_2
{}~({{\beta}_{ij}}^{+} - {{\beta}_{ij}}^{-})({\eta}_1) ~
D({\eta}_1-{\eta}_2) ~({{\beta}^{ij}}^{+} + {{\beta}^{ij}}^{-})
(\eta_2)\nn \\
& + & i~\int^\eta d\eta_1\int\limits_0^{\eta_1}d\eta_2
{}~ ({{\beta}_{ij}}^{+} - {{\beta}_{ij}}^{-})({\eta}_1)
N(\eta_1-\eta_2) ~({{\beta}^{ij}}^{+} - {{\beta}^{ij}}^{-})(\eta_2)
\tea
where $\Delta V(\eta) = \gamma_{even}(\eta) - \gamma_{odd}(\eta)
sgn(\eta)$ , $N(\eta) = 2{\tilde{\gamma}}(\eta)$ and $D(\eta) =
-2 \gamma_{odd}(\eta)$.
The first  two terms contribute a non-local potential to the effective
action but  do not contribute to the mixing of $\bpl$ and $\bmn$
histories like the third and fourth terms. We will show that the
third term with the kernel $D$ that is odd in the time domain
contributes to the dissipation and  the last term is associated
with noise.

\subsection{Noise}

Let us first concentrate on the fourth term. Its contribution to the
decoherence functional is given by
\be
exp [-~\int^\eta d\eta_1\int\limits_0^{\eta_1}d\eta_2
{}~ ({{\beta}_{ij}}^{+} - {{\beta}_{ij}}^{-})({\eta}_1)
N(\eta_1-\eta_2) ~({{\beta}^{ij}}^{+} - {{\beta}^{ij}}^{-})
(\eta_2)] \label{expnoise}
\te
This term can be rewritten using the following functional Gaussian identity
which states that the above expression  is equal to
\be
\int  D\xi(\eta) {\cal P}[\xi]exp [ i\int\limits_0^\eta
d{\eta'} \xi({\eta'})~ ({{\beta}_{ij}}^{+} - {{\beta}_{ij}}^{-})({\eta'})]
\te
where
\be
{\cal P}[\xi] = P_0 exp[{-                                \label{noisedist}
}\int\limits_0^\eta d{\eta}_1\int\limits_0^{\eta}d{\eta}_2
\ha \xi(\eta_1)N^{-1}
(\eta_1 - \eta_2)\xi(\eta_2)]
\te
is the functional distribution of $\xi(\eta)$ and $P_0$ is a
normalization factor given by
\be
{P_0}^{-1} = \int D\xi(\eta)exp [- \int\limits_0^\eta d\eta_1
\int\limits_0^\eta d\eta_2\xi(\eta_1) N^{-1}(\eta_1-\eta_2) \xi(\eta_2)].
\te
The influence functional can then be written as
\bea
e^{i\Gamma} &=& \int D\xi(\eta){\cal P}[\xi] exp{i\bar{\Gamma}
[\bpl, \bmn ,\xi]}\nn \\
&\equiv & {< exp {i\bar{\Gamma}
[\bpl, \bmn ,\xi]}  >}_{\xi}
\tea
where the angled brackets denote an average with respect to the
stochastic distribution ${\cal P}[\xi]$.
The modified influence action $\bar{\Gamma}[ \bpl, \bmn, \xi]$
is given by
\bea
\bar{\Gamma}[ \bpl, \bmn, \xi] & = &
  ~\int^{\eta}\int^{\eta}
d{\eta}_1d{\eta}_2~{\beta_{ij}}^{+} ({\eta}_1)\Delta V({\eta}_1
-{\eta}_2){\beta^{ij}}^{+}({\eta}_2)\nn \\
& - & ~\int^{\eta}\int^{\eta}
d{\eta}_1 d{\eta}_2~{\beta_{ij}}^{-} ({\eta}_1)\Delta V({\eta}_1
-{\eta}_2){\beta^{ij}}^{-}({\eta}_2)\nn \\
& - & ~\int\limits_0^{\eta}d{\eta}_1\int\limits_0^{{\eta}_1} d{\eta}_2
{}~({{\beta}_{ij}}^{+} - {{\beta}_{ij}}^{-})({\eta}_1) ~
D({\eta}_1-{\eta}_2) ~({{\beta}^{ij}}^{+} + {{\beta}^{ij}}^{-})
(\eta_2)\nn \\
&-& ~\int d\eta' \xi(\eta'){{\beta}_{ij}}^{+}
   +  ~\int d\eta' \xi(\eta'){{\beta}_{ij}}^{-}
\tea
The  term coupling a stochastic source $\xi$ to $\beta$ will manifest itself as
the noise in the equation of motion derived from this effective action. We see
that the influence action $\Gamma$ can be  written as an average of
$\bar{\Gamma}$ over this stochastic distribution function.

The decoherence functional can thus also be written as a stochastic average
\be
{\cal D}[ a^{+},{\beta}^{+}; a^{-},{\beta}^{-}]
 =  <e^{i \hat S_{eff} (a^{+},
{\beta}^{+}; a^{-},
{\beta}^{-};\xi)}>_\xi
\te
where the full effective action $\hat S_{eff}$ is given by
\bea
\hat S_{eff} &=&
S_{g}[a^{+}, {\beta}^{+}] + ~\int d\eta' \xi(\eta'){{\beta}_{ij}}^{+}
- \{S_{g}[a^{-}, {\beta}^{-}] + ~\int d\eta' \xi(\eta'){{\beta}_{ij}}^{-}\}
\nn \\
& - & ~\int\limits_0^{\eta}d{\eta}_1\int\limits_0^{{\eta}_1} d{\eta}_2
{}~({{\beta}_{ij}}^{+} - {{\beta}_{ij}}^{-})({\eta}_1) ~
D({\eta}_1-{\eta}_2) ~({{\beta}^{ij}}^{+} + {{\beta}^{ij}}^{-})
(\eta_2) \label{seff}
\tea
Our relevant equations of motion will be derived by varying $\hat S_{eff}$.
{}From  this equation we can view $\xi(\eta)$ as an external stochastic
force linearly coupled to $\beta$, though the linearity is a feature
specific to truncation of the perturbation series at quadratic order in the
effective action. In general we will have non-linear coupling.

Since the distribution functional (\ref{noisedist}) is Gaussian, this is a
Gaussian type noise, which is completely characterized by
\bea
{<\xi(\eta)>}_{\xi} &=& 0 \nn\\
{<\xi(\eta_1)\xi(\eta_2)>}_\xi &=& N(\eta_1-\eta_2)
\tea
Therefore the non-local kernel $N(\eta_1-\eta_2)$ is just the two-point
time-correlation function of the external stochastic source
$\xi(\eta)$. Since this correlation function is non-local, this noise is
colored.  As suggested in \cite{HuPhysica,HMLA}
we believe this is a rather general feature of noise of cosmological origin.\

\subsection{Einstein-Langevin Equation and Fluctuation-Dissipation Relation}

We will now show how this noise can be incorporated into the equation of
motion as a Langevin type equation. In this process we will also
demonstrate the role of the kernel $D$ in providing dissipation. The
key difference from the earlier treatment \cite{CH87} is that
the equation of motion will be derived from  the quantity
$\hat S_{eff}( \beta_+ , \beta_- , \xi)$ rather than the ``noise
averaged" quantity ${S_{eff}( \beta_+ , \beta_- )}$.
This has also been discussed in other contexts in \cite{HPZ2,nfsg,HM3}.
The first step is to write $\hat S_{eff}( \beta_+ , \beta_- , \xi)$
in terms of the following variables
\bea
\bar\beta_{ij} &=& {1\over 2} ( \bp + \bm) \nn \\
\Delta             &=& \bp - \bm
\tea
The equation of motion is then derived as
\be
{\delta \hat S_{eff}( \bar \beta_{ij} , \Delta)\over \delta \Delta }
{\Big | }_{ a^+ = a^- = a} = 0
\te
yielding
\bea
& & -2 \kappa{d\over d\eta} ( a^2 \bb') + {1\over 30{(4\pi)}^2}
{d^2\over d{\eta}^2} [\bb''ln(\mu a)]  + {1\over 90{(4\pi)}^2} {d\over d\eta}
\left\{ \left[ {\Big({a'\over a}\Big)}^2 + \Big({a''\over a}\Big)\right]
\bb''\right\}\nn \\
& + & \int d{\eta}_1 \gamma_{ren} (\eta - \eta_1)\bb(\eta_1)
= - j_{ij}(\eta) + {\xi}_{ij}(\eta)             \label{eom}
\tea
Here $j_{ij}$ is an external source term added in order to switch on
the anisotropy in the distant past \cite{HarHu}.
It is worth comparing these results with those in \cite{CH87} where
similar equations were deduced from the CTP effective action.
Comparing Eq. (\ref{eom}) with Eq. (3.18) in \cite{CH87} we find
that they are exactly the same except for the stochastic force ${\xi}_{ij}$
on the right hand side.
The real and causal kernel $K_4$ there (including the
numerical factor $1/[30(4\pi)]^2$) is identical to our kernel
$\gamma_{ren}$ . We will show that the odd part of this
kernel can be associated with dissipation.
One could  in fact interpret Eq.(3.18) obtained by Calzetta and
Hu as Eq. (\ref{eom}) averaged with respect to the noise
distribution. Since this is a Gaussian noise, $<\xi> = 0 $,
we obtain Eq. (3.18) of \cite{CH87}, where the
$\beta$'s are also to be interpreted as noise-averaged variables.
In this sense, we have gone beyond previous analysis
in  extracting the underlying stochastic behavior
that is lost in the smoothed out average version given in \cite{CH87}.

To make the analogy with a Langevin Equation more explicit it is convenient
to integrate (\ref{eom}) once with respect to $\eta$. This gives the following
equation

\bea
& & -2 \kappa a^2 \bb' + {1\over 30{(4\pi)}^2}
{d\over d{\eta}} [\bb''ln(\mu a)]  + {1\over 90{(4\pi)}^2}
\left\{ \left[ {\Big({a'\over a}\Big)}^2 + \Big({a''\over a}\Big)\right]
\bb''\right\}\nn \\
& + & \int d{\eta}_2\int d{\eta}_1 \gamma_{ren} (\eta_2 - \eta_1)\bb(\eta_1)
 = c_{ij} + s_{ij}
\tea
where $c_{ij}(\eta) = -\int d{\eta}' j_{ij}({\eta}')$ and
$s_{ij}(\eta) = \int d{\eta}' {\xi}_{ij}({\eta}')$.

Defining the variable $q_{ij} = d{\bb}/ d\eta$ we can write the above
equation  in the following  form
\be
{d\over d\eta}(M {dq_{ij}\over d\eta}) + {\cal K} {dq_{ij}\over d\eta}
+ k q_{ij} = c_{ij} + s_{ij}      \label{1int}
\te
where
\bea
M &=& {1\over 30{(4\pi)}^2}ln(\mu a) \\
k &=&  -2 \kappa a^2   + {1\over 90{(4\pi)}^2}
\left[ {\Big({a'\over a}\Big)}^2 + \Big({a''\over a}\Big)\right] \\
{\cal K}q_{ij} &=& \int d{\eta}_2\int d{\eta}_1 f(\eta_2 - \eta_1)
                   {dq_{ij}\over {d\eta_1}}
\tea
and $ d^{2}f(\eta)/ d{\eta}^2 = \gamma_{ren}$. This equation
is identical in form to the equation (3.15) in \cite{HuPhysica}
except for the term $s_{ij}$ on the right hand side, which is
indeed the stochastic contribution  from the noise anticipated there.
This equation is therefore a cosmological example of the Einstein-Langevin
equation for semiclassical gravity. It is in the form of a generalized
damped harmonic oscillator driven by a stochastic force $s_{ij}$.
(Of course the generalized mass $M$ and spring constant $k$ are time
dependent, so strictly speaking it has the damped harmonic oscillator
analogy only when these quantities are positive, as was also pointed
out in \cite{Paz90}.)

The second term on the left hand side of Eq. (\ref{1int})
represents the damping term involving a non-local (velocity dependent)
friction force. That this term is associated with dissipation can
be quickly seen as follows \cite{CH89}. In the Fourier transformed version
of a damped harmonic  oscillator equation the imaginary term
is associated with dissipation. Writing Eq. (\ref{1int}) in terms of the
Fourier transform $q_{ij}(\omega) = \int d\eta e^{-i\omega \eta} q_{ij}(\eta)$
we notice that the only imaginary contribution comes from the
second term on the left hand side, which can be written as
\be
F(q) = \int{d\omega\over 2\pi} e^{i\omega\eta}
{\gamma_{ren}(\omega)\over {\omega}^2} q_{ij}(\omega)
\te
where $\gamma_{ren}(\omega)$ is the Fourier transform of $\gamma_{ren}(\eta)$
defined in Eq. (\ref{odev}). Thus we see that the dissipation is associated
with the imaginary part of $\gamma_{ren}(\omega)$ or equivalently with
the odd part of the kernel $\gamma_{ren}(\eta)$ given by
$\gamma_{odd}$ defined in (\ref{odd}) as asserted before.
In fact, as in \cite{CH87,CH89} we can isolate the generalized (frequency
dependent) viscosity function $v(\omega)$ by writing
\be
iv(\omega)\omega q_{ij}(\omega) =
i {\rm Im } \gamma_{ren}(\omega)q_{ij}(\omega)
\te
{}From Eq. (\ref{odev}) we can identify $v(\omega)$ as
\be
v(\omega) = {{|\omega|}^3\over 60(4 {\pi}^2)}
\te
It can be shown that the kernel $\gamma_{ren}$ is related to particle
production
\cite{nfsg,HuSin}.
Now that we have identified the noise and dissipation kernels
$N(\eta)$ and $\gamma_{odd}(\eta)$ respectively, we can
go ahead and write down the fluctuation-dissipation relation in
analogy  with the quantum Brownian model \cite{HPZ2,HuBelgium}.
Defining
\be
D (\eta) = - 2 \gamma_{odd}(\eta) = {d\over d\eta}\gamma(\eta)
\te
The fluctuation-dissipation relation has the familiar form
\be
N(\eta) = \int\limits_0^\infty d{\eta}' K(\eta-{\eta}')\gamma({\eta}')
\label{fdan}
\te
where the FD kernel $K(\eta)$ is given by
\be
K(\eta) = \int\limits_0^{\infty} {d\omega\over \pi}~\omega \cos\omega \eta
\te
This proves the conjecture of \cite{HuPhysica} that there exists
a fluctuation-dissipation relation for the description of the backreaction
effect of particle creation in cosmological spacetimes.
We see that the FD kernel is identical with that appearing in the
fluctuation-dissipation relation  obtained in other cases with different
system-environment couplings \cite{HPZ2}, but with the environment at
zero temperature as we have here.  Hence this also vindicates
the previous observation  \cite{HPZ2,Zhang,SinSor} that the zero temperature
fluctuation-dissipation relation is insensitive  to the nature of the
system-bath coupling. Since we have not taken the bath at a finite
temperature, thermal fluctuations play no role in the above relation
and it summarizes the effect solely of quantum fluctuations. Effect of
thermal fluctuations can be included easily and we expect a FDR to hold
for finite temperature particle creation and backreaction as well.

\subsection{Related Problems}

In the problem studied in Part II the system is itself a field. By contrast,
the problem discussed here  has only a few degrees of freedom
interacting with a field.
I will end this section with a
description of related problems of interest which can be described by a model
similar to this.\\

1) To begin with, one can study the nature of quantum noise
of a quantum field in a cosmological or black hole spacetime. The system can
be a simple harmonic oscillator. It can act as a particle detector in
a black hole or de Sitter universe, and with the statistical method
one can derive the Hawking radiation as a result of the magnification
of quantum noise without invoking any geometric concept.
This was carried out recently by
Matacz and I \cite{HMLA,HM2}, and mentioned as point 3c) in the Outline.
One can also use a harmonic oscillator with negative kinetic energy to
depict the dynamics of a Robertson-Walker universe, and the oscillator-field
model to describe  quantum field processes in curved spacetimes. This was
discussed in \cite{HM3}.

2) For particle creation-backreaction problems similar to the Bianchi-I model
studied here, one can extend the present results to that of, say, a minimally-
coupled scalar field in a Robertson-Walker or de Sitter universe.
This mimics the linearized graviton modes and has practical use for the
description of primordial stochastic gravitons. The particle production
problem was first studied by Grishchuk \cite{Gri}, the backreaction by
Grishchuk \cite{Gri76} and  Hu and Parker \cite{HuPar77} via canonical
quantization methods, and by Hartle \cite{Har81}
and Calzetta and Hu \cite{CH87} via the in-out and in-in effective action
method respectively. The influence functional method expounded here
would enable one to get the noise, which is related to the fluctuations
in graviton number \cite{nfsg,HuSin}, and derive the Einstein-Langevin
equation for the study of graviton production and metric fluctuations.
Calzetta and I are presently working on this problem.

3) Sciama \cite{Sciama} was the one who had the great insight of
seeing the Hawking and
Unruh radiation as an excitation of the vacuum fluctuations and the detector
response as following a dissipation-fluctuation relation. Inspired by this
way of thinking, I posed for myself the challenge of showing a fluctuation-
dissipation relation at work for quantum fields in a general cosmological
spacetime, not required to possess an event horizon (see Mottola's work
on de Sitter universe \cite{Mottola}) and use such general relations
in quantum open systems to understand  the particle creation-backreaction
problem. The work with Calzetta \cite{CH87} and Sinha \cite{HuSin}
completed this quest and opens the door for providing
a statistical mechanical interpretation
of these processes, in terms of the relation of quantum and thermal
fluctuations
of quantum fields when subjected to various conditions of spacetime curvature,
topology and dynamics. The Einstein- Langevin equation in these cases  will
also provide insight into the behavior of metric fluctuations and the
relation of quantum gravity to semiclassical gravity as a phase transition
process.

4) The extension of the present and related models to quantum cosmology
is of interest. The backreaction of `higher' gravitational perturbation
modes on the homogeneous background spacetime can be formulated in
the framework of quantum cosmology \cite{HuErice} and be used to address the
validity of the minisuperspace approximation, as was done by Sinha and I
\cite{SinHu}. Doing the same extension from CTP to IF, one can derive
the noise associated with the truncated inhomogeneous cosmological modes.
One can also define an entropy function from the reduced density matrices,
which measures the information loss in the minisuperspace truncation.
These can perhaps be called geometrodynamic noise and gravitational entropy.
It would be interesting to compare this statistical mechanical
definition with the definition suggested by Penrose \cite{Penrose} in classical
general relativity and by me in the semiclassical context \cite{Hu83}.
I have noted some initial thoughts on this problem
at the Waseda conference \cite{HuWaseda}. Details are to be found in
\cite{HuSinSTN}. \\

{\bf Acknowledgement}
I thank the organizers of this conference, especially Randy Kobes and Gabor
Kunstatter, for making this nice meeting possible and for their warm
hospitality.
I  take great pleasure also to express my appreciation to  my former associates
and students  who have contributed greatly to this research program,
Esteban Calzetta, Andrew Matacz, Juan Pablo Paz
Sukanya Sinha, and Yuhong Zhang for their lively discussions and enjoyable
collaborations over many years. Research is supported
in part by the National Science Foundation under grant PHY91-19726.

\end{document}